\RequirePackage[2018-12-01]{latexrelease}
\documentclass{rQUF2e}

\usepackage{epstopdf}
\usepackage{subfigure}

\theoremstyle{plain}

\theoremstyle{definition}

\theoremstyle{remark}

\begin{document}


\title{An Empirical Analysis on Financial Markets: Insights from the Application of Statistical Physics}

\author{Haochen Li$\dag$\thanks{haochen\_li@kcl.ac.uk, yi.cao@ed.ac.uk, maria.polukarov@kcl.ac.uk}, Yi Cao${\ddag}$, Maria Polukarov$\dag$, Carmine Ventre$^{\ast}$${\dag}$\thanks{$^\ast$Corresponding author: carmine.ventre@kcl.ac.uk}\\
\affil{$\dag$King's College London, London, WC2R 2LS, UK\\ $\ddag$The University of Edinburgh, Edinburgh, EH8 9YL, UK} \received{August 2023} }

\maketitle

\begin{abstract}
In this study, we introduce a physical model inspired by statistical physics for predicting price volatility and expected returns by leveraging Level 3 order book data. By drawing parallels between orders in the limit order book and particles in a physical system, we establish unique measures for the system's kinetic energy and momentum as a way to comprehend and evaluate the state of limit order book. Our model goes beyond examining merely the top layers of the order book by introducing the concept of `active depth', a computationally-efficient approach for identifying order book levels that have impact on price dynamics. We empirically demonstrate that our model outperforms the benchmarks of traditional approaches and machine learning algorithm. Our model provides a nuanced comprehension of market microstructure and produces more accurate forecasts on volatility and expected returns. By incorporating principles of statistical physics, this research offers valuable insights on understanding the behaviours of market participants and order book dynamics.
\end{abstract}


\begin{classcode}C51, C53, C55, C58, G17\end{classcode}

\section{Introduction}

In this article, we carry out an empirical analysis of employing a novel model rooted in the principles of statistical physics for limit order book. We do so by predicting perhaps the most important measures in financial markets, that of price volatility and changes. This study distinguishes itself with three significant contributions. 

{\bf Implication on asset pricing.} First, we establish novel benchmarks for the predictive accuracy of cryptocurrency asset price volatility and expected return utilising Level 3 order book data. The accuracy of our model is justified by high predictive R-squared values and prediction accuracy compared to previous literature, illustrating its robustness across different market conditions. Our model bypasses the need for training samples or parameter fitting since it takes full advantage of the market microstructure information inherently enclosed in the Level 3 order book data. \cite{challet2003non}, \cite{eisler2012price} and \cite{cont2014price} studied the price impact of market order submissions, and limit order submissions and cancellations that contained in the Level 3 order book data, and discussed how these action might affect the limit order book (LOB). This information is adequate for estimating the impact of both current and forthcoming orders on the order book dynamics and, subsequently, price dynamics. We underline significant economic implications for investors leveraging these predictive measures, which display a considerable edge over conventional techniques such as the Volume-Synchronized Probability of Informed Trading (VPIN) and the Order Flow Imbalance (OFI) when predicting volatility and expected asset returns.

Volatility and return prediction carry tangible economic significance. The principal objective of asset pricing revolves around comprehending the patterns of expected return and its accompanying volatility, as presented by \cite{french1987expected}. Even if expected returns were observed completely, the requirement for theories explaining their behaviour still persists and the same for the empirical analysis validating them. However, the tasks of measuring volatility and return are well-known intricate: at a lower frequency time scale (like hourly or daily), return variations are primarily influenced by unpredictable news events that impact prices. Upon examination of identical datasets, distinct volatility estimators can manifest disparate characteristics. Notably, according to \cite{cont2014price}, these estimators exhibit varied intra-day patterns in different underlying assets. On the other hand, at high-frequency trading scales, they are shadowed by the market microstructure noise generated by market participants, including market makers, informed traders, and noise traders.

\cite{cont2014price} proved that the order book depth deeper than best bid and best ask contain information relevant to the price dynamics. Based on this empirical finding, a recent strand of literature attempted machine learning methods to capture the volatility and expected return of assets. \cite{doering2017convolutional} developed convolutional neural networks (CNN) to predict price and its volatility. \cite{kercheval2015modelling} employed Support Vector Machine (SVM) on the order book data. \cite{sirignano2019deep} and \cite{sirignano2019universal} developed a deep Long Short-Term Memory (LSTM) model to study the snapshot data of the order book. However, due to the restriction of computing power and model complexity of machine learning models, they are not feasible to study the information in deeper depths of the order book. Our proposed physics-based model defines the concept of Active Depth to distinguish the order book depths relevant with a simplicity in model complexity. The model interpretability is another shortcoming of the machine learning methods labeled as `black boxes' because of the difficulty to understand the precise reasoning behind them, while the physics-based model provides a straightforward intuition to explain the order book dynamics by modelling it as a physical system.

Market microstructure noise is another important topic to consider as it captures various of frictions intrinsic to the trading process. As argued by \cite{ait2008high}, these noises include the bid-ask bounces, discreteness of price fluctuations, differences in trade sizes, the informational content of price changes, strategies of the order flow, and inventory management effects. Eliminating the impact of these noises during investigating the asset price is meaningful for the asset pricing practices, particularly in the high-frequency trading markets.

We provide a novel technique to model the order book dynamics that is possible to bypass hence eliminate the impact of both market news and market microstructure noises as they arise from the time series of asset price, volume, bid-ask spread or any other explanatory variables generated by order book dynamics. As a result, our research acts as a foundation upon which future studies can be built, moving the field towards a more comprehensive understanding of the mechanisms that dominate asset pricing and, by extension, the broader financial market dynamics.

{\bf A physical model on limit order book.} Second, we incorporate the empirical asset pricing field with the physical models, which offers a unique perspective on market dynamics. While asset pricing traditionally focuses on determining the intrinsic value of assets, an understanding of the market microstructure – the system through which the trading of assets takes place – is crucial to comprehend how the supply and demand for assets are matched and how prices are ultimately discovered. 

\cite{bouchaud2002statistical} along with \cite{potters2003more} elucidated the statistical intricacies inherent to the order book. These empirical investigations served as the foundation of 'zero-intelligence' models. Such models posited that the 'stylized facts' are an emergent outcome of order flow properties and the intrinsic structure of the order book, without the need for the 'rationality' assumptions. In their studies on limit order book, \cite{bouchaud2002statistical} and \cite{potters2003more} offered a systematic exploration of empirical properties, and developed models based on statistical physics that linked the financial markets with physical systems.

The increasing interest in applying physical models to finance, both in academia and in the industry, has yielded empirical studies in the field of asset pricing and the field of Econophysics. \cite{bouchaud2003fluctuations} examined the complex interplay between order flow and price fluctuations, drawing analogies with the external stimuli in physical systems. Their insights unveiled the market `randomness'. \cite{farmer2004origin} through rigorous analysis, highlighted a connection between power-law tails in price changes and order flow dynamics, aligning with phenomena observed in physics. \cite{farmer2006market} explored the effects of supply-demand on price movements, contrasting with the Efficient Market Hypothesis. Their findings drew parallels with long-range correlations in physical systems. \cite{borland2012statistical} looked into the statistical pattern during market panics, by harnessing the concept of self-organizing systems from physics. Inspired from these studies in Econophysics, our analysis aim to provide a detailed investigation of the utility of a physical model applied to the challenging problems of modelling the LOB.

\cite{bak1997price} offered a distinctive perspective on price dynamics within a multifaceted agent stock market. Drawing parallels with physical systems, orders were conceptualised as particles on a price trajectory, where intersections resulted in transactions. \cite{yura:2014financial} conceptualised orders as fluid particles, using an interaction range defined by order activity to segment the order book into layers. \cite{yura:2015financial} expanded this model, linking price movement velocity to changes in the inner layer's time-averaged limit orders, and introduced a financial Knudsen number. 

Approaching the complex structure of the limit order book from a statistical physics perspective allows us to view the orders as interactive elements, and the price fluctuation of an asset as a macroscopic property of this complex system. The historical price data merely encapsulates past alterations in this macroscopic property, analogous to the temperature in a fluid system. Nonetheless, examining the evolution of a fluid system necessitates knowledge of inherent physical parameters such as density, mass, and velocity. These parameters cannot be seized from historical temperature data alone; they require an understanding of the complex system. The key of the argument lies in the fact that the limit order book system is vastly more information-rich compared to historical price data.

The current literatures of asset pricing does not address to this point but merely studies the time series of asset price, while statistical physics provides the framework and techniques to build such models. By treating the limit order book as a statistical physical system, we can tap into a deeper, more macroscopic level of information. This approach, akin to the methods employed in statistical physics, allows us to better understand the underlying mechanisms that drive the dynamics of asset prices. By recognizing that price movements are a macroscopic feature of a more complex system of interacting orders, we can gain new insights into the asset pricing process. This perspective opens up novel avenues for research and presents opportunities for the development of more robust models for understanding the price discovery and predicting asset prices.

We employ the term `physical model' to refer to a novel method that mimics the dynamics of the order activities. We introduce a statistical physics perspective to model the limit order book, drawing an analogy between financial markets and physics. It constructs a microscopic model of the orders, conceptualising them as particles within the complex system of the limit order book. Order submissions are interpreted as particles entering the system, order cancellations as particles exiting the system, and each order transaction as a particle annihilation. This allows us to calculate the change in ‘kinetic energy’ and ‘momentum’ within the system and estimate the impact of each activity. Our method considers the different quoted prices of each single order upon submission and cancellation, estimates the impact of each order activity on the order book, and utilises the microscopic dynamics of order book activities to construct a systemic description of market behaviour. We evaluate their efficacy in predicting price changes in the Bitcoin and LUNA markets. Our findings suggest that these new measures offer enhanced predictive power and capacity to accurately assess the market behaviour, price volatility and return of the cryptocurrencies, outperforming existing models such as the VPIN and OFI measure. This approach also enables us to visually highlight events occurring within the high-frequency trading market. Based on the current literature, few analytical tools harness the vast amount of data in this domain to effectively illustrate financial market phenomena.

In our analysis, we find that the predictive power of these measures, when analyzed through the lens of our physical model, surpasses that of the benchmark techniques. Our findings show an improvement in the ability to forecast price volatility and return in the asset. Moreover, the physical model's success in addressing the inefficiencies of traditional prediction models leads us to our next empirical fact: the use of physical models to analyse market microstructure is viable when complemented with accurate description of the order book complex system measured by the physical models. This empirical research serves to justify the role of such models in the broader architecture of the financial markets, which offers a significant advantage in capturing the complexity of the order book dynamics that the traditional methods often fail to accomplish.

However, it's important to note that while our physical model significantly enhances our understanding of asset returns, it is fundamentally a measurement tool. It merely explains the process that the limit and market orders discover the price and impact its dynamics, while it does not provide insights into the economic mechanisms that drive these activities. As such, while the physical model can significantly describe market behaviour and enhance prediction, it must be complemented with economic theories that investigate the underlying mechanisms driving asset returns. This remains an exciting direction for future research, as there is an emerging literature that combines these advanced analytical tools with economic theories to provide a more comprehensive understanding of asset returns.

{\bf Active Depth in the order book.} While top layers of the order book play a pivotal role in short-term price determination, deeper levels encapsulate broader market sentiment and strategic placements, thereby holding crucial information for more long-term price predictions.

\cite{bouchaud2002statistical} emphasised the importance of understanding the full depth of order books and not just the top layers. Their models, which built on statistical physics, revealed the intricate patterns and correlations across different layers, underscoring the significance of considering deeper levels when analysing price dynamics. \cite{zovko2001power} highlighted the nuanced `patience' traders demonstrate, suggesting that order placement is not just concentrated near the best bid and ask, instead, many traders placed orders at deeper levels and the distribution decayed as a power law. Through simulation analysis, \cite{smith2003statistical} provided a statistical framework to describe the continuous double auction mechanism. Their theory encapsulated the significance of interactions at varying depths in the order book. By illustrating how orders at different layers influence price determination, they accentuated the importance of understanding the full spectrum of the order book for accurate price prediction. \cite{chiarella2002simulation} brought forth insights into the microstructure dynamics of double auction markets. Their agent-based model, with traders influenced by factors like chartists, fundamentalists, and noise, underscored that the depth of the order book was not static but evolved based on these influences. They advocate for an in-depth examination of the order book, suggesting that the deeper levels can provide substantial insights into forthcoming market movements.

However, due to the restriction of computing power and model complexity, it is impossible to study the information in all the depths of the order book. The market makers, informed traders, noise traders and all other market participants participates the market by submitting and cancelling the orders. The order book dynamics is impacted by these order activities. As a result, it is possible to find the depths that are encountering such activities. Based on this inspiration, we proposed a method to define and calculate the concept of active depth, which distinguish the depths that are active and contains the relevant information to impact the order book.

In conclusion, the physical model, with its purely empirical approach based on limit order book data, provides a fresh perspective that transcends the traditional assumptions of empirical and behavioral finance. This methodology offers a more nuanced and comprehensive analysis of market dynamics, heralding a new direction in financial market research. However, it is important to acknowledge that while this approach provides valuable insights, it does not entirely replace the need for understanding the underlying economic mechanisms or equilibrium. Complementary use of empirical, behavioral, and the physical model may provide a more holistic perspective of financial markets and asset pricing.

The structure of this article is as follows. The review on the current related literature is given in Section~\ref{Literature review}. In Section~\ref{Data}, we overview the source and format of the cryptocurrency order book data we used for our experiments. A physical model to describe the limit order book and the order activity is presented in Section~\ref{methodology}, as well as the definition and calculation of the pre-requisite concept called active depth. In Section~\ref{Energy} the kinetic energy measure from the physical model is compared to the VPIN method on predicting the volatility. In Section~\ref{Momentum} another measure of momentum is discussed in a comparative analysis to the Order Flow Imbalance method. At last, in Section~\ref{Benchmark}, a deep LSTM model is used as the benchmark model to carry out comparative analysis between it and the physics-based model.

\section{Literature Review}
\label{Literature review}
\noindent

We here give an overview of existing financial models and methodologies, focusing on those relevant to model the financial market with physics and those predicting price volatility and impact.

\cite{lillo2003master} conducted data analysis on the determinants of the price impact of individual transactions, accounting for factors such as volume and underlying stock characteristics. To standardise across different liquidity levels, they re-scaled the mean price change and volume, resulting in a uniform price-impact curve across various stocks. Their findings suggested a universal statistical law governing fluctuations from supply-and-demand equilibrium across different financial assets. They postulated the efficacy of modelling human activity as stochastic in certain contexts over rational models. Building on these insights, \cite{daniels2003quantitative} modeled the price formation process, assuming random order placement and cancellation, which allowed them to predict price movements successfully.

\cite{farmer2005predictive} further refined the modelling of order arrival, price movement, and implied supply-demand relationships in a continuous double auction context. They established functional relationships between order-arrival rates and market statistical characteristics, as well as price movements and order flow. Upon testing this model with trading data from different stocks, they found a shared price-impact curve, suggesting a universal statistical law. \cite{lillo2005key} demonstrated that the multidimensional stochastic process of bid-ask spread exhibits specific statistical attributes, and that fluctuations in the state of the limit order book drive price movements.

The aforementioned studies provide evidence that the state of the limit order book and movement of limit orders follow stochastic processes, revealing certain statistical laws governing the limit order book. Consequently, it becomes viable to study the rules governing order activities using statistical physics.

\cite{yura:2014financial} studied the LOB by drawing a parallel between financial market dynamics and Brownian motion. They segmented the order book into layers based on an `interaction range' defined by order activity. Consequently, they modeled the bid-ask spread and the nearest depths with the highest temporary volume as a colloidal financial Brownian particle. Their work provided valuable insights on investigating how order book structures drove price dynamics with statistical physics. \cite{li2023detecting} modelled the orders as particles and calculated the LOB system's momentum in the defined passive area, which they use to recognise the anomalous patterns and detect market manipulation activities. They showcased the potential of physical models on Level 3 order book data which may not be discernible using conventional financial econometrics tools. Although this work did not study the volatility and expected return, \cite{li2023detecting} offered an innovative way of viewing the LOB and analysing the behaviours of market participants.

\cite{easley2011microstructure} provided an in-depth analysis of the events of May 6, 2010, known as the `flash crash'. The authors argued that the crash was precipitated by automated execution of a large sell order in the E-mini contract, leading to two liquidity crises. They presented evidence that the liquidity problem was slowly developing in the hours and days before the collapse, with high and unbalanced volume but low liquidity. The authors also discussed the role of high-frequency trading (HFT) firms in this event, noting that these firms, which account for over 73\% of all U.S. equity trading volume, often act as market makers. The paper introduced a measure of order flow toxicity, VPIN, and demonstrated its predictive power for absolute returns and its potential use as a warning measure for the risk of a liquidity crash. The authors concluded that in markets dominated by high-frequency liquidity providers, unusually high levels of flow toxicity could lead to liquidity providers turning into liquidity consumers, potentially destabilising the market.

In the paper by \cite{andersen2014vpin}, the authors further critically examined the Volume-Synchronised Probability of Informed Trading (VPIN) and its variants. They argued against the significance of any specific metric unless it retained significance in formal tests that incorporate readily observable real-time market activity measures, such as trading intensity and implied volatility measures. The authors also discussed the volume clock, the Order Imbalance (OI) measure, and the Trade Classification (TR-VPIN) metric. They highlighted the complex interplay of parameters such as the volume bucket, the time bar, the trade classification indicator, and the length of the moving average in determining both the level and the dynamic behaviour of the metric. The authors concluded that these parameters were necessary to gauge the incremental information content of any new metric.

\cite{easley2021microstructure} further demonstrated the application of machine learning algorithms to predict and explain modern market microstructure phenomena. They investigated the efficacy of various microstructure measures, showing that they continued to provide insights into price dynamics in current complex markets. Interestingly, some microstructure features with high explanatory power exhibited low predictive power, and vice versa. The authors also found that some microstructure-based measures were useful for out-of-sample prediction of various market statistics, leading to questions about market efficiency. Their results were derived using 87 of the most liquid futures contracts across all asset classes. The authors suggested that as markets become faster and data more copious, the market microstructure played a critical role in predicting market behavior. They argued that despite the increased complexity of trading strategies and the empirical measures of microstructure, machine learning techniques can play an important role in the evolution of microstructure research.

Regarding to the market microstructure analysis based on the order flow, \cite{cont2014price} empirically investigated the influence of order book events on equity prices using NYSE TAQ data. It proposed a simplified model using a single variable, the Order Flow Imbalance (OFI), to linearly explain mid-price changes, demonstrating robustness across various stocks and timescales. Based on this work, \cite{shen2015order} presented a compelling exploration of the Volume Order Imbalance (VOI), which provided a robust framework for understanding market dynamics and offers valuable insights into the supply and demand analysis on market microstructure level. \cite{cartea2018enhancing} provided an in-depth exploration of how trading strategies can be improved by incorporating signals from the order book. Their research offered valuable insights into the application of high-frequency order book data in strategy optimization, thereby broadening the understanding of market microstructure and its implications for trading strategies. Their rigorous mathematical approach and insightful findings underscored the importance of incorporating granular market data to enhance trading efficacy.

In recent years, applying machine learning approaches on finance is an emerging topic while there are relatively few that using such approaches to study the LOB. \cite{nevmyvaka2006reinforcement} introduced machine learning to the LOB by applying reinforcement learning on optimal order execution. \cite{kercheval2015modelling} applied support vector machines (SVM) to predict the direction move of the mid-price for stocks. \cite{sirignano2019deep} argued that neural networks outperforms SVM as well as other machine learning methods for LOB as they are more capable of dealing with high-dimensional data and capturing nonlinear relationships, while SVM can only provide binary classification instead of indicate probabilities. \cite{sirignano2019universal} further developed a deep neural network (DNN) model with Long Short-Term Memory (LSTM) units. They trained an universal model with the LOB data of about 500 NASDAQ stocks to predict the direction move of price.

\section{Data}
\label{Data}
\noindent

This section details the data sets used in the study. It describes the source of the data, which includes transaction data for Bitcoin and LUNA, and the time frame covered. It also outlines the steps taken to clean and prepare the data for analysis.

Financial markets are typically organised around a central record, the limit order book (LOB), which logs all orders placed by traders. Our study aims to process Level 3 order book data and summarise the complete information about order activity and order book dynamics. Rather than relying on time-based snapshot data of the order book, we propose a more comprehensive approach that processes every detail and message of the order book data, thereby preserving all temporal information. 

Our research drew upon high-frequency order book data from the Coinbase exchange, collected via a websocket feed. We gathered Level 3 order book data for the BTC/USD and LUNA/USD cryptocurrency pairs. As the most granular market data, Level 3 data includes records of order submission, cancellation, and matching. While Level 1 data, on the other hand, is the most basic, offering only Open/High/Low/Close price and volume data for a specific period. And Level 2 data provides the price and aggregated volume in the depths beyond the bid-ask spread in the order book, where the bid-ask spread is the difference between the highest price a buyer is willing to pay (bid) and the lowest price a seller is willing to accept (ask) for a particular asset.

The submission of a limit buy order may create a buy order either beneath the best bid price, at the best bid price, or even surpass the bid-ask spread which will result in an immediate match transaction. The extent of its movement along the price axis is designated as the distance between its quoted price and the active bid price. A cancellation of a buy order is construed as the movement of the buy order from its original quoted price to the active bid price which will be defined in Section~\ref{Calculation of Active Depth}. If an order is received with a quoted price that is equal to or worse than an existing order on the other side(usually at best bid or best ask), it will be executed forthwith and produce a `match' record. If the order is not instantly executed or only partially executed, it will establish an `open' record in the order book and stay at a specific depth of the limit order book. In our dataset, the majority of open orders were cancelled, at an approximate ratio of 98.5\%.

We investigated the Level 3 order book data for the BTC/USD pair during 15:00-16:00 on 28/11/2022, and for the LUNA/USD pair during 02:00-03:00 on 12/05/2022, while the LUNA was facing the significant fall. The original datasets are composed of two datasets - full channel data and ticker channel data. The full channel encompasses all of the Level 3 order book data discussed previously, and is event-driven, with every order activity (submission, cancellation, and matches) leaving a record. On the other hand, the ticker channel contains snapshot information of the order book, including the best bid price and best ask price, with a new record in this channel created each time a match occurs. However, the aggregated quoted volume of the best bid and best ask price was available for the BTC/USD data, but not for the LUNA/USD data. There were more than 2,700,000 lines of records in the full channel data and 47,000 in the ticker channel for each one-hour BTC/USD dataset, while the numbers were about 480,000 and 35,000 for LUNA/USD. We also used the 1-second LOB snapshot data for the BTC/USD pair during the 12-hour span between 23:00 11/08/2023 - 11:00 12/08/2023 to train deep LSTM model, and this training set contained 43,200 samples of LOB snapshots. For the following period of 11:00-12:00 on 12/08/2023, the LOB snapshot data was used to test the deep LSTM model and to compare with the physics-based model that used the Level 3 order book data meantime.

The minimum price precision for both the BTC/USD and LUNA/USD pairs is 0.01 USD, and the minimum trading volume unit (order size) is 0.00001 BTC for the BTC/USD pair and 0.001 LUNA for the LUNA/USD pair. The minimum precision of the recorded timestamp is one microsecond (1e-6 second). A timestamp is recorded when an order activity (event) occurs, hence the timestamps of the event-driven records are discrete and non-consecutive.

For data pre-processing, we merged both data sets by retrieving the corresponding order book information that was updated in the most recent record from the ticker channel data, for each record in the full channel data. Specifically, we recorded the best bid, best ask prices when the last event of each $\Delta t=0.1$ time period occurred, treating these as the best bid and best ask prices for each order activity happening within the same sampling interval. Then, for the datasets of both BTC/USD and LUNA/USD pairs, we sampled these event-based records at a frequency of $\Delta t=0.1$ second, in preparation for applying our proposed model to compute and aggregate the physical measures.

\section{Methodology}
\label{methodology}

\subsection{The Physical Model}
\label{The Physical Model}
\noindent

In this section, we will propose a model for the LOB inspired by concepts from physics, which models the LOB as a physical system where the conservation of momentum and energy can be applied. The denotations of prices, order velocities and momentum are built upon the foundations established in our previous work \cite{li2023detecting}. We will extend these concepts to introduce and define the novel measures such as the active depth, total volume of reacted order, and kinetic energy of the LOB.

In the LOB, the best bid and best ask prices are represented by $b_M$ and $a_M$, and the market midprice is subsequently denoted by $z_M=\frac{b_M+a_M}{2}$. Additionally, the market match price $p^*$ is noted whenever a trade execution occurs. For a given time $t$, the marke midprice and match price are denoted by $z_M(t)$ and $p^*(t)$ respectively. The quantity of limit orders at a specific quoted price $x$ is denoted by $N(x,t)$ which is invariably non-negative. Over the interval $[t-\Delta t, t]$, the \emph{velocity} of the match price, $v^*(t)$, is calculated by the rate of change in the match price with $v^*(t)=\frac{p^*(t)-p^*(t-\Delta t)}{\Delta t}$.

\begin{figure}
\centerline{\includegraphics[height=2.5in]{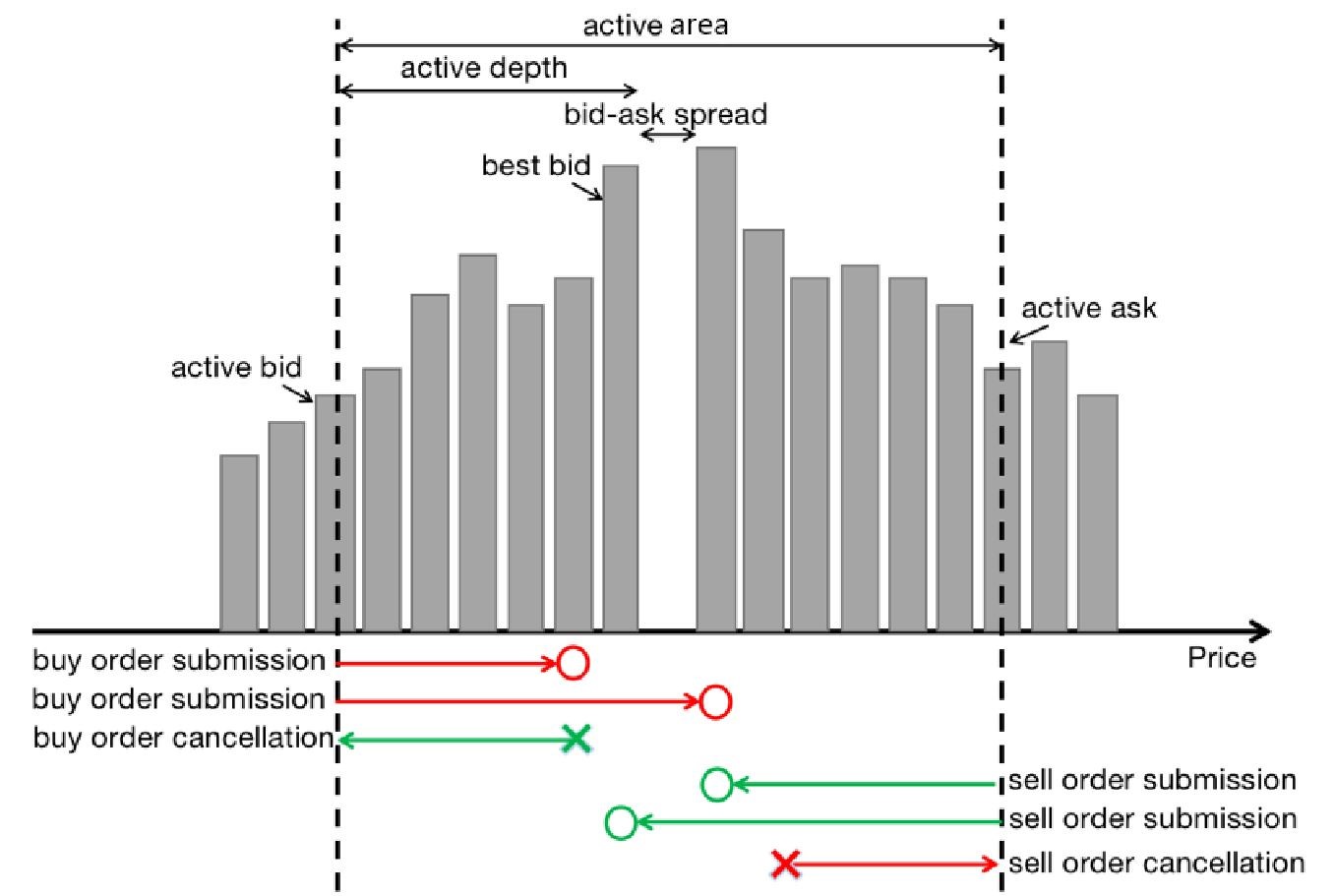}}
\caption{Schematic of the order book} \label{fig:foo2}
\end{figure}

It was observed that order activities predominantly cluster around the market midprice. A so-called \emph{active area} is identified as an extended range at a particular depth within the LOB, oriented around the bid-ask spread as illustrated in Figure~\ref{fig:foo2}. The dynamics of the active area closely mirror those of the market midprice; within this range, order submissions and cancellations exhibit elevated activity, while beyond this range, limit orders generally remain stationary. Consequently, the \emph{active bid price} and \emph{active ask price} are derived as the best bid and best ask prices adjusted by the active depth defined in Section~\ref{Calculation of Active Depth}.

In this study, orders are conceptualised as physical particles moving on a one-dimensional axis. The order size corresponds to the particle mass, while its displacement is akin to the distance of particle movement. Over a sampling period $\Delta t$, the velocity $v(t)$ of an order is computed as the quotient of its displacement $\Delta p$ and the elapsed time $\Delta t$. The order velocity is only defined within the active area, bounded by $[b_M-\alpha, a_M+\alpha]$ where $\alpha$ is the active depth which will be defined in Section~\ref{Calculation of Active Depth}.

Over each sampling period $(T-\Delta t,T]$, we define the submission activity of a limit buy order with quoted price $p^*(t)$ at a specific at time $t$ as its movement from the active bid price $b_M-\alpha$ to its quoted price $p^*(t)$ during this time interval. For a limit buy order of size $s$ submitted to its quoted price $p^*(t)$ at time $t$ during the sampling period $(T-\Delta t,T]$, $t \in (T-\Delta t,T]$. As a result, the order's velocity can be computed with 
\begin{equation}
    v(t) = \frac{p^*(t)-(b_M(T-\Delta t)-\alpha)}{\Delta t}.
\label{eq:2}
\end{equation}
It is noteworthy that for limit buy orders exceeding the best ask price $a_M$, the effective quoted price becomes $p^*(t)=a_M$ at time $T - \Delta t$. 

In contrast, a canceled buy order is considered to move from its initial quoted price $p^*(t)$ to the active bid price, $b_M-\alpha$ at time $T - \Delta t$. Its velocity is thus calculated as:
\begin{equation}
    v(t) = \frac{(b_M(T-\Delta t)-\alpha)-p^*(t)}{\Delta t}.
\label{eq:3}
\end{equation}
Similarly, for a newly submitted limit sell order, the velocity is
\begin{equation}
    v(t) = \frac{p^*(t)-(a_M(T-\Delta t)+\alpha)}{\Delta t}.
\label{eq:4}
\end{equation}
Here, if the quoted price is below the best bid price $b_M$, the effective quoted price becomes $p^*(t)=b_M$ at time $T - \Delta t$. For canceled sell orders, the velocity is computed as:
\begin{equation}
    v(t) = \frac{(a_M(T-\Delta t)+\alpha)-p^*(t)}{\Delta t}.
\label{eq:5}
\end{equation}

The depth of the order book will be raised to specify the location of limit orders with regard to the inside of the order book (best bid price for buy orders and best ask price for sell orders) at the same side. At time $t$, the depth of a specific limit order quoted at price $p^*$ is denoted by $\gamma(p^*,t)$ and defined by
\begin{equation}
    \gamma(p^*,t)=[b_M(T-\Delta t)-p^*]
\end{equation}
for the bid side, where the buy orders are normally quoted at price lower than the best bid price. And
\begin{equation}
    \gamma(p^*,t)=[p^*-a_M(T-\Delta t)]
\end{equation}
for the ask side, where the sell orders are normally quoted at price higher than the best ask price. Note that the depth for a limit order could be negative value in the case that it is quoted between the bid-ask spread.

Let $N_\gamma(t)$ denote the total number of limit orders submitted at depth $\gamma$ of the LOB during time $t$, and $N'_\gamma(t)$ for the total number of limit orders cancelled at depth $\gamma$ at time $t$. We define the total volume of reacted orders (either cancelled or submitted) inside depth $\gamma$ during $T-\Delta t$ to $T$ as

\begin{equation}
    \Delta V(\gamma)=\sum_{t \in (T-\Delta t,T]} \sum_{\gamma^* = b_M-\gamma}^{a_M+\gamma}(\sum_{i=0}^{N_\gamma(t)}s_i + \sum_{j=0}^{N'_\gamma(t)}s'_j)
\label{eq:total volume of reacted order}
\end{equation}
where $s_i$ is the order size of the $i^{th}$ submitted limit order quoted at this depth, while $s'_j$ is the order size of the $j^{th}$ cancelled limit order quoted at this depth.

Now, we can define the kinetic energy and momentum of an order akin the physical kinetic energy and momentum of an object. That is, the product of half its size and squared velocity $\frac{1}{2}s\cdot v^2$ for kinetic energy, and the product of its size and velocity $s\cdot v$ for momentum. Thus, to calculate the kinetic energy and momentum of submitted market and limit orders we shall use equations (\ref{eq:2}) and (\ref{eq:4}) for velocity, and for cancelled limit orders -- equations (\ref{eq:3}) and (\ref{eq:5}). 

Recall the denotation of $N_\gamma (t)$ for the total number of limit orders at depth $\gamma$ of the LOB at time $t$. We then calculate the cumulative sum of kinetic energy and momentum respectively at time $t$ by aggregating the kinetic energy and momentum resulting from all order activities occurred in the order book depths $\gamma$ ranging from the lower bound to the upper bound of the active area. Specifically, the kinetic energy of the order book at time $t$ is 
\begin{equation}\label{eq:active area kinetic energy}
    E(t) = \sum_{t \in (0,t]} \sum_{\gamma = b_M-\alpha}^{a_M+\alpha} ( \sum_{i=0}^{N_\gamma(t)}\frac{1}{2}s_i\cdot {v_i}^2 + \sum_{j=0}^{N'_\gamma(t)}\frac{1}{2}s'_j\cdot {v'_j}^2)
\end{equation}
and the momentum of the order book at time $t$ is
\begin{equation}\label{eq:active area momentum}
    P(t) = \sum_{t \in (0,t]} \sum_{\gamma = b_M-\alpha}^{a_M+\alpha} (\sum_{i=0}^{N_\gamma(t)}s_i\cdot v_i + \sum_{j=0}^{N'_\gamma(t)}s'_j\cdot v'_j)
\end{equation}
where $s_i$ and $v_i$ are the order size and velocity of the $i^{th}$ submitted limit order quoted at this depth, while $s'_j$ and $v'_j$ are the order size and velocity of the $j^{th}$ cancelled limit order quoted at this depth.

Importantly, our approach models the order book as a complex system, and emphasizes the vicinity surrounding the bid-ask spread, where order activities exhibit the highest degree of activity, and assigns greater weight to orders quoted in closer proximity to the bid-ask spread as they exert a more substantial influence on the bid-ask spread.

\subsection{Calculation of Active Depth}
\label{Calculation of Active Depth}
\noindent

This section explains the methodology used to calculate the active range, a crucial component in the proposed physical model. We will present the details how to establish the definition and calculation on the upper and lower boundaries of the active range.

In the preceding section, we introduced the concept of the active area, defined by the order book depth $\alpha$, wherein the majority of order activity takes place. This section will focus on the computational methodology for deriving this pivotal depth and elaborate on the empirical analysis supporting its efficacy.

To extract this depth empirically and determine the active area $\alpha$, we scrutinize the interconnection between the orders proximate to the bid-ask spread and their response in terms of order cancellation or submission, subject to movements in the match price. Our approach employs a one-second time window to compute the change in the total volume of reacted orders $\Delta V(\gamma)$ within different order book depth $\gamma$ using equation~\ref{eq:total volume of reacted order}. Concurrently, we assess the corresponding distance traversed by the match price $v^*(t)$ within this identical time frame.

\begin{figure}
\begin{center}
\begin{minipage}{160mm}
\subfigure[BTC/USD market.]{
\resizebox*{8cm}{!}{\includegraphics{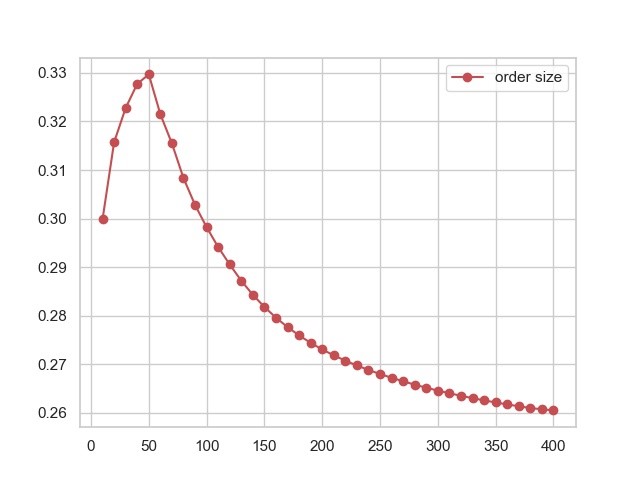}}\label{fig:BTC0420_00calc}}
\subfigure[LUNA/USD market]{
\resizebox*{8cm}{!}{\includegraphics{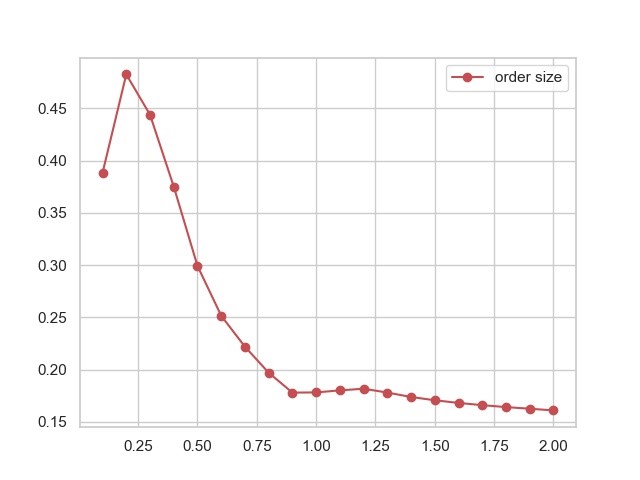}}\label{fig:LUNA0511_17calc}}
\caption{Correlation analysis of the match price and the change in the total volume of reacted orders.
\label{fig:calc}}
\end{minipage}
\end{center}
\end{figure}

Our analysis further extends to study the cross-correlation function $\text{Corr}[v^*(T), \Delta V(\gamma)]$ to investigate the relationship between changes in the match price and alterations in the total volume of orders reacted at depth $\gamma$. The graphical representation of this relationship is presented in Figure~\ref{fig:calc}.

A close examination of Figure~\ref{fig:BTC0420_00calc} reveals the correlation coefficient peaking at depth 50 for the BTC/USD market. In contrast, for the LUNA/USD market, the maximum correlation is observed at a considerably shallower depth of 0.2, as illustrated in Figure~\ref{fig:LUNA0511_17calc}. This is due to the different proportions of tick size (minimum price increment unit) to the price of the respective underlying asset. For Bitcoin, this ratio is approximately \$0.01 to about \$16,000, whereas for LUNA, it's significantly smaller, at \$0.01 to about \$2. These graphical illustrations portray a quantifiable correlation between fluctuations in the match price and corresponding changes in the total volume of reacted orders, as a function of the order book depth.

This characteristic pattern demonstrates a collective trend following behaviour of orders, thereby substantiating our active area concept. In particular, at an order book depth of 50 for Bitcoin and 0.2 for LUNA, the high-frequency traders exhibit a pronounced tendency to modify the largest volume of order in response to the movement of the match price. 

Now we can define the active depth $\alpha$ by:

\begin{equation}\label{eq:active depth}
    \alpha = \operatorname*{arg\,max}_\gamma \text{Corr}[v^*(T), \Delta V(\gamma)]
\end{equation}

This empirical evidence amplifies the pivotal role of the active area in the financial market dynamics and underscores its utility in assessing the market microstructure. It offers a valuable lens through which to understand and anticipate the behavior of high-frequency traders. The findings contribute to the broader understanding of market liquidity and depth, and could have important implications for market participants and regulators alike.

\section{Empirical Findings}
\label{Empirical Findings}
\noindent

In this section we present the results of applying the proposed physical model to the datasets of Bitcoin and LUNA. We report the outcomes of their analysis and interpret these findings, highlighting the ability of our physical model to describe the order book dynamics and market behaviour.

In the physical world, it is not feasible to study the movements of a specific particle. Somehow, if we consider a massive number of particles as a whole complex system, we may discover the statistical rules of their collective behaviour and motion. It is similar in the order book dynamics, as the orders are placed by individual traders according to their supply and demand needs, which drive irregular market oscillation and noise. However, when we gather all of the order records and information, some collective behaviour emerges. From these collective behaviours of the orders, we may have a glance at the market behaviour and hence uncover the intrinsic reasons leading to the order book dynamics and price dynamics.

We evaluate our framework in the context of the May 2022 flash crash of the LUNA cryptocurrency. By examining the LOB data corresponding to the LUNA/USD pair, we observe how `limit' and `market' orders exerted opposing forces on the market. Limit orders, being less aggressive, set a `limit' price for transactions, while market orders, which are more aggressive, buy or sell at the current `market' price. During the LUNA flash crash, market orders attempted to raise the price unsuccessfully, while limit orders consistently engaged in selling activities. This suggests that the decline in LUNA's price was primarily due to the dominance of limit orders, which effectively counterbalanced the strength of market orders. This implies that the flash crash was largely triggered by traders submitting limit orders. Our methodology also provides a means of visualising order activities, thereby enabling academics, industry participants, and regulatory authorities to gain a more comprehensive understanding of the market. This facilitates a more effective comprehension of market behaviour and market microstructure.

\begin{figure}[!ht]
\begin{center}
\includegraphics[height=2.5in]{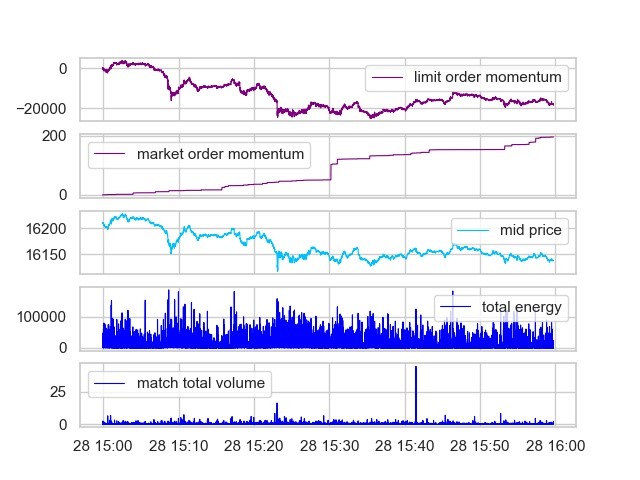}
\end{center}
\caption{Measures on BTC/USD market}
\label{fig:BTC_measure}
\end{figure}

\begin{figure}[!ht]
\begin{center}
\includegraphics[height=2.5in]{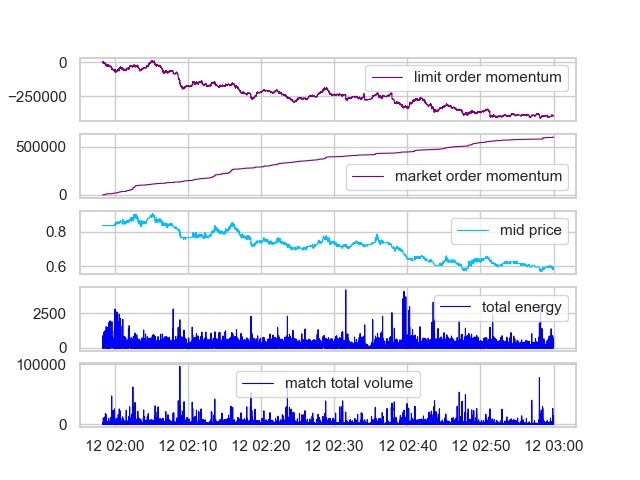}
\end{center}
\caption{Measures on LUNA/USD market}
\label{fig:LUNA_measure}
\end{figure}

For BTC/USD pair with active depth of 50, after separating the momentum for the limit orders and the market orders, we have results in Figure~\ref{fig:BTC_measure} where the first row of the plot is the cumulative sum of net momentum for the limit orders and the second row is for the market orders. We can see that there is a jump in the cumulative sum of net momentum for the market orders. This was brought by the placement of a huge market buy order. Besides, traders submitting market orders are significantly less than limit orders which is relected by the value of momentum for limit orders and market orders. Similar experiment on the LUNA/USD pair with active depth of 0.2 lead to the results in Figure~\ref{fig:LUNA_measure}. 

An interesting phenomenon of market microstructure can be observed that the market orders were constantly trying to push the price upwards (failing to do so) while the limit orders were selling all the time.

In general, the kinetic energy is a scalar that indicates the extent of order activities, a.k.a, when the LOB have higher energy, the more order activities are taking place and the price dynamic is facing larger fluctuation. On the other hand, the momentum is a vector that implies the direction that the limit orders and market orders are making effort to push.

Our model adopts the principle of kinetic energy and momentum, originating from the realm of physics, to signify the force or impact of each order to the LOB system. Through the computation of kinetic energy and momentum encompassing all order activities, such as submissions and cancellations, this technique sheds light on market behaviour, liquidity, and other market microstructure characteristics. Moreover, the technique facilitates an in-depth, microscale examination of the order book by employing Level 3 order book data. This degree of granularity allows for a more precise analysis on order activities.

\section{Comparison with Other Microstructure Measures}
\label{Other Measures}
\noindent

\subsection{The Roll Measure}

The Roll measure, introduced by~\cite{roll1984simple}, is a financial metric used to estimate the bid-ask spread and the associated liquidity of a financial asset. A larger Roll measure indicates a wider bid-ask spread and, consequently, lower liquidity and higher transaction costs. Conversely, a smaller Roll measure suggests a narrower bid-ask spread, higher liquidity, and lower transaction costs. The Roll measure provides an aggregated perspective based on price changes and is based on the efficient market hypothesis, which assumes that prices fully reflect all available information.

Our physics-based model, on the other hand, offers a microscopic view of the market, modeling orders as physical particles and calculating momentum based on Level 3 order book data. Unlike the Roll measure, this model does not assume market efficiency, focusing instead on statistically describing the state of the order book and market behaviour. The physical model provides a comprehensive analysis of the order book, captures the impact of high-frequency trading on market microstructure, and offers intuitive visualizations to identify the market behaviour. Moreover, this model is flexible and adaptable, allowing for its extension to incorporate additional physical concepts or analyze the other asset classes.

Comparatively, the physics-based model and the Roll measure offer different insights into financial markets. While the Roll measure provides a simple and effective estimator of market liquidity based on price data, the physical model provides a more detailed view of market microstructure, and offers a comprehensive analysis of the order book data.

The physics-based model utilises Level 3 order book data, which includes every order activity (limit orders, cancellations, and match) and their respective information. This rich dataset allows for a deeper understanding of market microstructure, as it provides a complete picture of market participants' actions and intentions, unlike the Roll measure, which only relies on transaction price data.

In conclusion, while both the physical model and the Roll measure contribute valuable insights into market dynamics, the physical model's ability to provide a comprehensive, detailed view of market microstructure, detect market manipulation, and consider the impact of high-frequency trading presents a more nuanced and adaptable tool for researchers interested in understanding the complexities of modern financial markets. As financial markets continue to evolve and become more complex, it is essential for microstructure measures to adapt and capture these changes effectively, and the physical model seems well-positioned to meet this challenge.

\subsection{The Kyle's Lambda}

Kyle's Lambda is a measure of the market impact cost of trading a financial asset, introduced by~\cite{kyle1985continuous}. This measure quantifies the sensitivity of a financial asset's price to trading activity. The higher the lambda, the greater the market impact cost, indicating that the asset's price is highly sensitive to trading activity. Therefore, Kyle's Lambda is a critical tool to understand the potential costs associated with trading a particular financial instrument. By understanding the market impact cost, market participants can make more informed decisions about the timing, size, and execution strategy of their trades, thus minimizing the impact on the price and maximizing the profitability of their transactions. Kyle’s lambda can be interpreted as the cost of demanding a certain amount of liquidity over a given time period and is an inverse proxy for liquidity, with higher values of lambda implying lower liquidity and market depth.

The physics-based proposed in this article and Kyle's Lambda both contribute significantly to our understanding of the microstructure of financial markets. However, they represent different aspects and utilize distinct methodologies.

Kyle's Lambda provides a theoretical framework for understanding the relationship between trade size, price impact, and the level of asymmetric information in a market. It primarily aims to quantify the market impact cost associated with a specific trade size, thus providing a valuable tool for understanding liquidity provision in financial markets. Kyle's Lambda is estimated using regression analysis, with the change in price regressed on the trade size and other control variables. This method relies on historical data and assumes a particular functional form for the relationship between trade size and price impact.

On the other hand, the physics-based model proposed is developed using a particle system model of the order book, allowing for a granular examination of the dynamics of individual orders and their impact on the order book. The physical model is primarily designed to describe the order book dynamics and does not directly measure the cost of trading as Kyle's Lambda does. The new model, with its two measures, kinetic energy and momentum, focuses on the underlying impact exerted by orders on the order book. The physical model is versatile, applicable across various financial markets and instruments, and can be applied to real-time data, allowing for continuous monitoring of order book dynamics. In general, the physics-based model proves in its ability by providing a more direct and intuitive way to study market microstructure.

\subsection{The Amihud Measure}

The Amihud measure, a financial metric introduced by~\cite{amihud2002illiquidity}, evaluates the liquidity of a financial asset by gauging its price impact in relation to trading volume.

While both the physics-based model and the Amihud measure provide essential insights into market behaviors, their methodological foundations are distinct. The Amihud measure calculates liquidity by assessing the price impact of trading volume. In contrast, the physics-based model takes a unique approach of viewing orders as particles and computes their physical measures. Moreover, the physics-based model does not predicate on market efficiency, while the Amihud measure is founded on the relationship between price impact and trading volume, serving as a proxy for liquidity.

The two models also diverge significantly in their interpretations of market microstructure and their data requirements. The physics-based model offers a more nuanced perspective of the market, allowing for the identification of order book dynamics. Conversely, the Amihud measure provides a macroscopic viewpoint based on daily return and volume data. Further, while the physics-based model requires Level 3 order book data to garner detailed temporal information, the Amihud measure operates on daily return and volume data.

Despite their divergences, both models have their distinct strengths. However, the physics-based model presents advantages, making it more suitable for certain applications in empirical financial analysis. Firstly, by employing Level 3 order book data, it offers a comprehensive understanding of market microstructure, superior to the Amihud measure that relies on aggregated daily return and volume data. Secondly, the physics-based model can identify the market behaviour of limit and market orders. Moreover, the physics-based model demonstrates flexibility and adaptability, capable of being extended to include additional physical concepts or to study market manipulation across various asset classes. Finally, by considering the impact of high-frequency trading and analyzing the impact of individual orders, it captures the effects of HFT on market microstructure.

\section{Energy and VPIN on Volatility Prediction}
\label{Energy}
\noindent

In the realm of financial market analysis, predicting volatility is an ongoing challenge and a critical component for both risk management and trading strategies. Traditional volatility prediction models have often relied on measures like historical volatility, implied volatility, GARCH models, and more recently, VPIN (Volume-Synchronized Probability of Informed Trading). However, we proposed a novel approach to predict short-term volatility by employing the concept of Energy Change, inspired by physical systems.

The VPIN model, which measures the imbalanced volume of trades, is based on the idea that informed traders will trade more aggressively, leading to an imbalance in the trade volume and hence, increased volatility. The model has been popularly used for its ability to capture the arrival rate of informed traders and its subsequent impact on volatility. However, the VPIN model primarily focuses on transaction data, and as a result, might not fully capture the dynamics of the limit order book.

In contrast, the kinetic energy measure proposed offers a fresh perspective by drawing an analogy between physical systems and financial markets. The Energy measure is calculated as half the sum of the squared order velocity in the defined active range of the order book. This measure is expected to capture the degree of demand to produce a change for different order activities, thus evaluating a macroscopic view of the order book dynamics.

\subsection{Introduction on the VPIN Method}

The Volume-Synchronized Probability of Informed Trading (VPIN) represents a cutting-edge analytical tool, garnering increasing interest within the realm of market microstructure studies. This tool serves as a metric for the disequilibrium between informed and uninformed market trades. The cornerstone of VPIN lies in the hypothesis that informed traders leverage private information, thus provoking an imbalance in trading volumes.

To derive VPIN, one divides the trade volume into discrete segments, or `buckets', and subsequently juxtaposes the volume of buyer-oriented trades to that of seller-initiated trades within each segment. Normalizing the disparity between these two volumes by the total volume yields the VPIN. A high VPIN value signals a heightened likelihood of informed trading, indicative of a substantial disequilibrium between purchase and sale orders. In contrast, a low VPIN suggests a harmonious market characterized by equivalent levels of purchasing and selling.

In the sphere of financial markets, VPIN serves as an instantaneous indicator of market states, offering crucial insights into the degree of information asymmetry in the market. This asymmetry could presage periods of intense volatility. Hence, the VPIN may serve as an instrument for forecasting short-term market fluctuations, proving particularly beneficial for high-frequency traders and market makers.

Abrupt VPIN fluctuations, particularly sharp escalations, may serve as an early warning system for impending market volatility. Some empirical studies suggest that elevated VPIN values can anticipate price volatility driven by liquidity and market imbalances. Consequently, traders and financial institutions frequently employ VPIN as a risk management instrument. However, it is crucial to bear in mind that while VPIN is a valuable metric, it does carry its set of limitations. The VPIN model presumes that all informed trades will instigate price variations, an assumption that may not always hold. Moreover, the selection of volume bucket size can significantly influence the VPIN output and there lacks a universal method choosing the optimal bucket size.

\subsection{Historical Volatility}

In the context of time series analysis, the Granger causality test is a statistical hypothesis test for determining whether one time series is useful in forecasting another and provide evidence of predictive power.

\begin{figure}
\begin{center}
\begin{minipage}{160mm}
\subfigure[BTC/USD market.]{
\resizebox*{8cm}{!}{\includegraphics{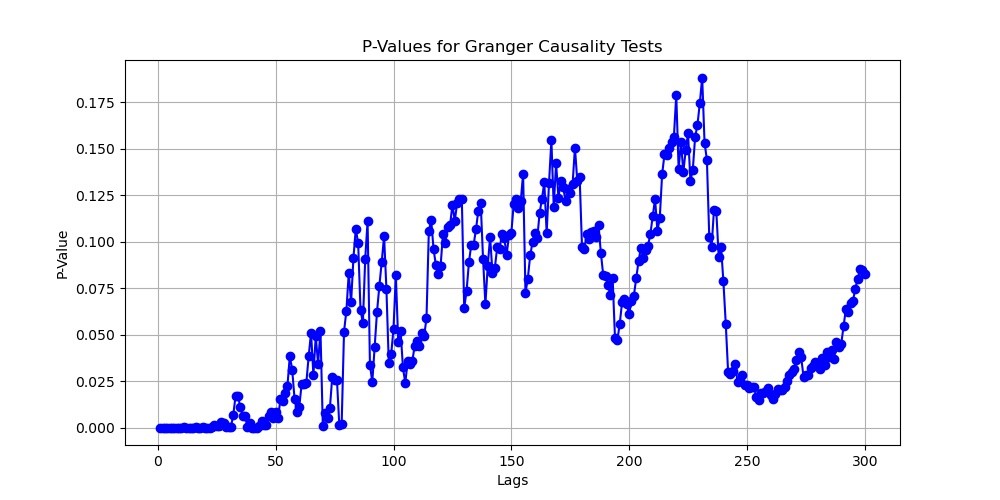}}\label{fig:BTC_granger_vpin}}
\subfigure[LUNA/USD market]{
\resizebox*{8cm}{!}{\includegraphics{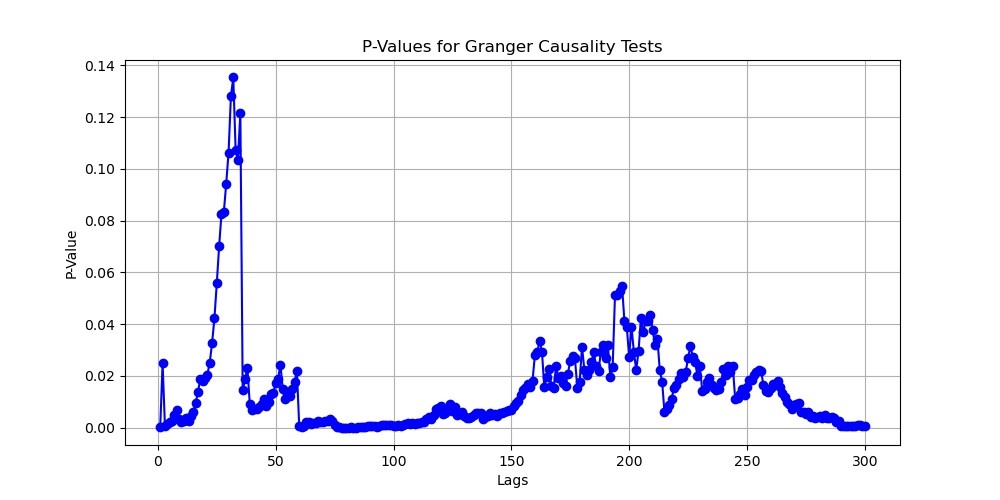}}\label{fig:LUNA_granger_vpin}}
\caption{Granger causality test on the 5-minute volatility against the VPIN.
\label{fig:granger_vpin}}
\end{minipage}
\end{center}
\end{figure}

\begin{figure}
\begin{center}
\begin{minipage}{160mm}
\subfigure[BTC/USD market.]{
\resizebox*{8cm}{!}{\includegraphics{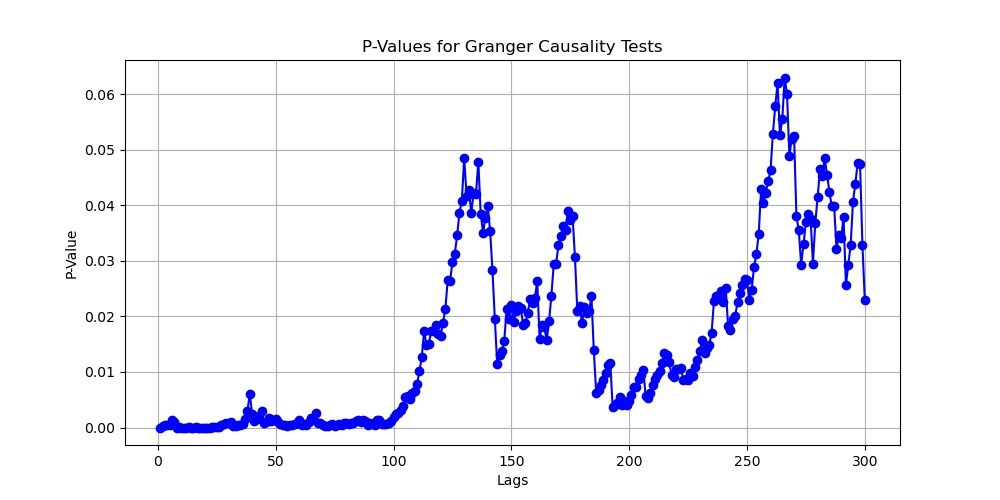}}\label{fig:BTC_granger_energy}}
\subfigure[LUNA/USD market]{
\resizebox*{8cm}{!}{\includegraphics{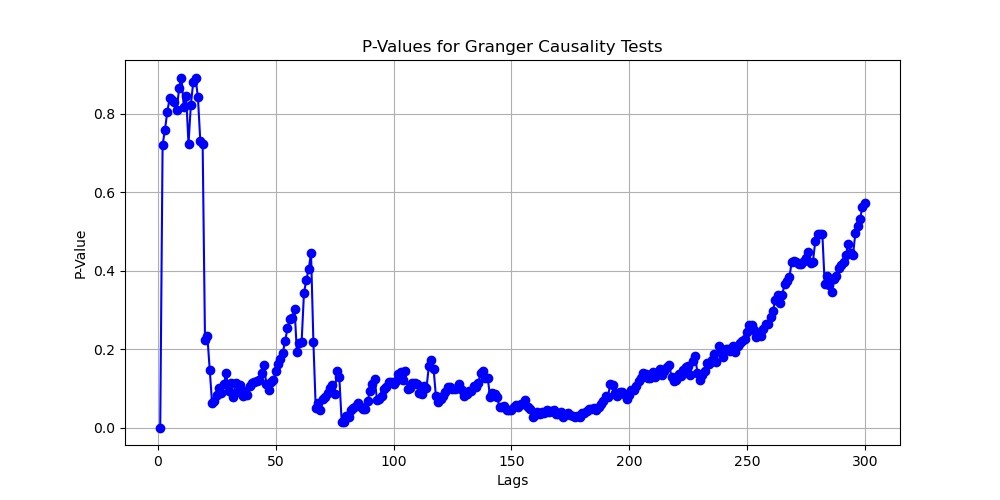}}\label{fig:LUNA_granger_energy}}
\caption{Granger causality test on the 5-minute volatility against the kinetic energy.
\label{fig:granger_energy}}
\end{minipage}
\end{center}
\end{figure}

The results of P-values in the Granger causality test on the 5-minute volatility against the VPIN time series lagging from 1 to 300 seconds for Bitcoin and LUNA are shown in Figure~\ref{fig:granger_vpin}. The similar results on the 5-minute volatility against the kinetic energy are shown in Figure~\ref{fig:granger_energy}.

A small p-value indicates strong evidence against the null hypothesis and suggests that the latter time series Granger-cause the former time series. The VPIN data performs the best in predicting the 5-minute volatility that is about 0-50 seconds behinds for Bitcoin and about 55-150 seconds for LUNA. And the kinetic energy performs the best predicting the 5-minute volatility about 0-100 seconds behinds for Bitcoin while it acts terribly in predicting the 5-minute volatility for LUNA.

In general, we can conclude that the VPIN of LUNA performs better than its kinetic energy in predicting the 5-minute volatility. while the kinetic energy of Bitcoin outperform its VPIN.

\subsection{Local Volatility}

As we are studying the high-frequency trading data, we adopt a different method to calculate the instantaneous volatility. This new method seeks to address a potential shortcoming of the traditional standard deviation calculation in the context of financial asset price data, particularly when the asset price exhibits a trending behavior.

The traditional calculation of volatility involves computing the standard deviation of returns, which is based on the mean return. However, when the asset price exhibits a trending behavior, the mean return could be significantly different from zero, and changing the length of the time window used for the calculation could result in artificially high volatility values. This is because more returns would be further away from the mean, leading to a larger sum of squared deviations.

To address this issue, our Local Volatility measure computes the standard deviation of the differences between consecutive asset prices, rather than their returns. This provides a measure of how much the price changes from one tick to the next, irrespective of the overall trend.

The Local Volatility for an asset with a price series $P_t$ is defined as:

\begin{equation}
    \sigma_{LV} = \sqrt{\frac{\sum_{t=1}^{T}(P_{t}-P_{t-1})^2}{T}}
\end{equation}
where $P_t$ represents the price at time $t$, and $T$ is the total number of price observations, the symbol $\sigma_{LV}$ represents the Local Volatility.

This equation calculates the differences between consecutive prices ($P_{t}-P_{t-1})^2$, sums them up, divides by the total number of observations to calculate the mean squared difference, and finally takes the square root to obtain the standard deviation of price changes. This Local Volatility measure provides a more meaningful estimate of the instantaneous volatility of an underlying asset, particularly when the asset price exhibits a trending behavior.

\begin{figure}
\begin{center}
\begin{minipage}{160mm}
\subfigure[BTC/USD market.]{
\resizebox*{8cm}{!}{\includegraphics{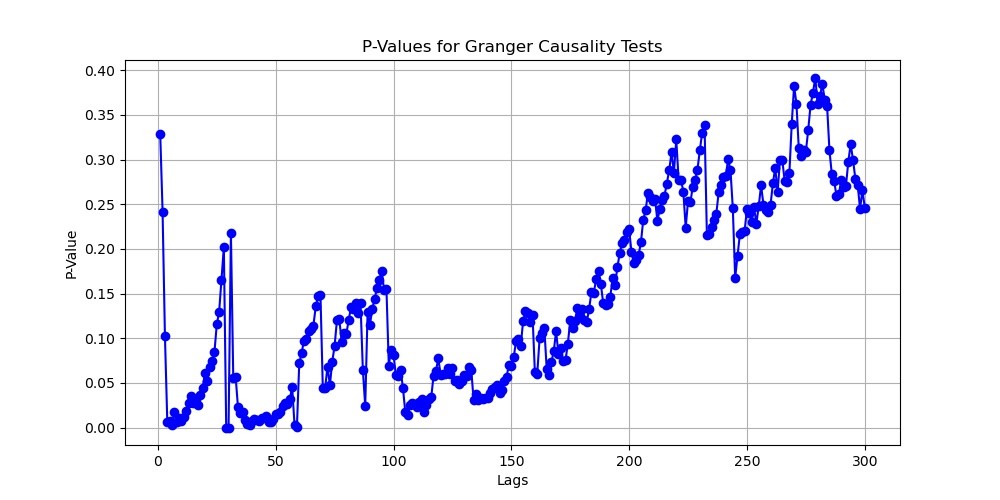}}\label{fig:BTC_granger_vpin2}}
\subfigure[LUNA/USD market]{
\resizebox*{8cm}{!}{\includegraphics{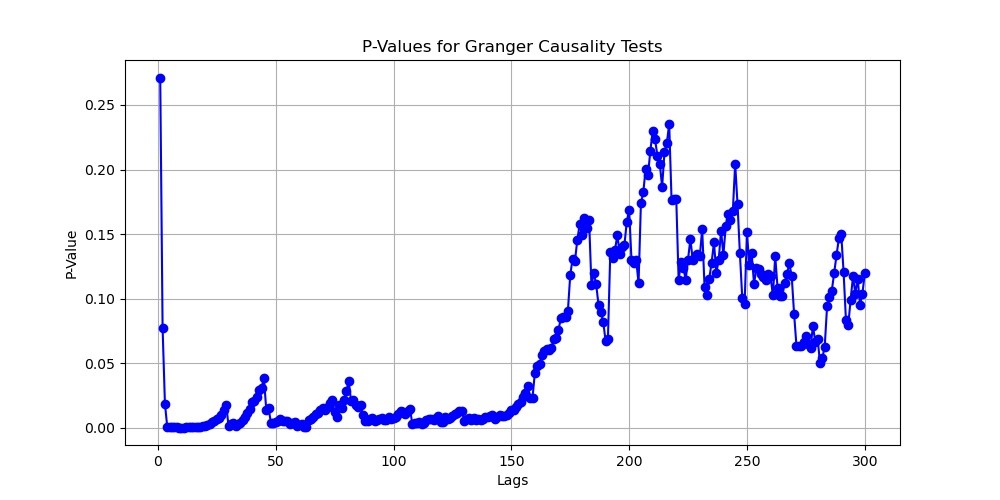}}\label{fig:LUNA_granger_vpin2}}
\caption{Granger causality test on the 30-second Local Volatility against the VPIN.
\label{fig:granger_vpin2}}
\end{minipage}
\end{center}
\end{figure}

\begin{figure}
\begin{center}
\begin{minipage}{160mm}
\subfigure[BTC/USD market.]{
\resizebox*{8cm}{!}{\includegraphics{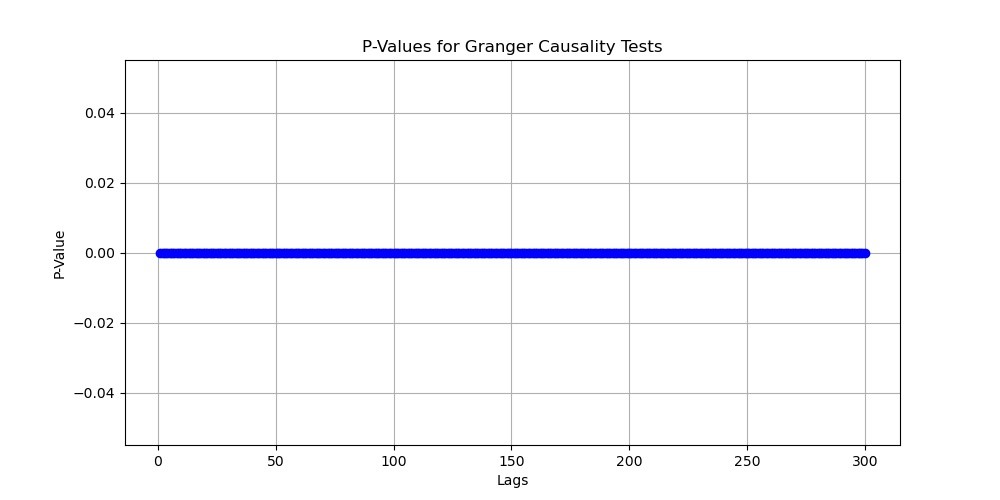}}\label{fig:BTC_granger_energy2}}
\subfigure[LUNA/USD market]{
\resizebox*{8cm}{!}{\includegraphics{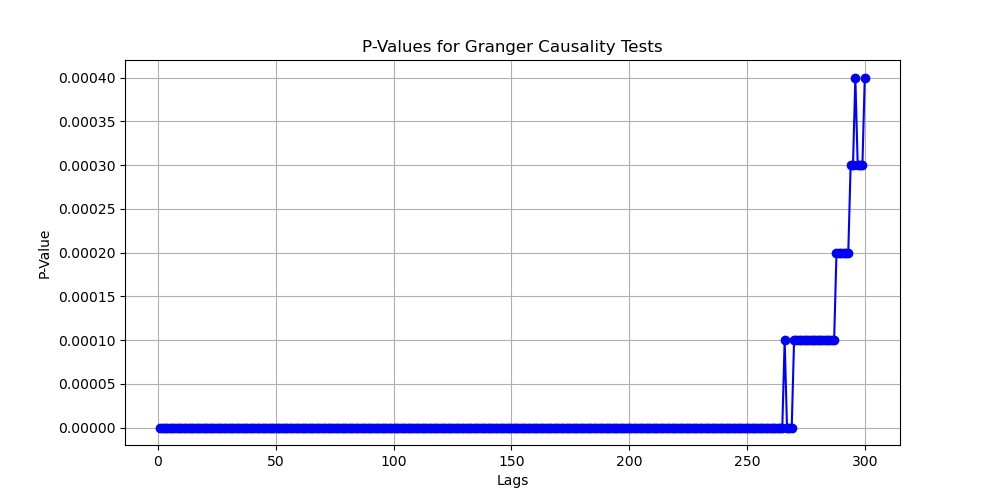}}\label{fig:LUNA_granger_energy2}}
\caption{Granger causality test on the 30-second Local Volatility against the kinetic energy.
\label{fig:granger_energy2}}
\end{minipage}
\end{center}
\end{figure}

Regarding to the calculated 30-second Local Volatility, we carried out the Granger causality test again, and the results of 30-second Local Volatility against the VPIN are shown in Figure~\ref{fig:granger_vpin2} while that against the kinetic energy in Figure~\ref{fig:granger_energy2}.

It is apparent that the VPIN method has a much worse performance in the instantaneous volatility data with higher sampling frequency. And these results prove that the kinetic energy owns significantly advantage compared to the VPIN method in predicting the instantaneous volatility in a high-frequency trading market.

\subsection{Comparative Analysis}

The empirical analysis reveals that the kinetic energy measure outperforms the VPIN model in predicting short-term volatility for both Bitcoin and LUNA. This superior performance to the kinetic energy measure's ability to capture not only the effects of individual order activity but also the overall order book dynamics. Our model includes the LOB depth by computing the kinetic energy measure of each order activity according to the LOB depth of their quoted price so that their different impact on the order book can be analysed quantitatively, which are often overlooked by models focusing solely on transaction data.

The findings suggest that while the VPIN model has its merits, the kinetic energy measure provides a more encompassing view of order book dynamics and is thus a more effective tool for predicting short-term volatility. This highlights the potential benefits of integrating concepts from statistical physics into financial markets.

These results underscore the importance of considering multiple facets of market data when predicting volatility. The kinetic energy measure's success points to the potential value of developing hybrid models that combine transaction data with other elements of the LOB. Such an approach could provide a more accurate and comprehensive understanding of the complex and rapidly changing dynamics that characterize financial markets.

In conclusion, the comparative analysis of the kinetic energy measure and the VPIN model represents a significant contribution to the literature on volatility prediction. It not only provides a novel perspective on volatility prediction but also sparks a potential shift in the methodological paradigm from purely econometric models to an interdisciplinary approach combining financial economics and statistical physics. Further research exploring this interdisciplinary approach could open up new vistas for improving volatility prediction in financial markets.

\section{Momentum and Order Flow Imbalance on Price Prediction}
\label{Momentum}
\noindent

Forecasting price changes in financial markets forms the foundation of trading strategies and risk management paradigms. Conventional methodologies frequently depend on the Order Flow Imbalance (OFI), a metric that characterizes shifts in supply and demand. We introduced a novel measure - the physical momentum, conceptually rooted in physical systems - for anticipating near-term price fluctuations.

The OFI has been widely recognized and utilised in microstructure studies, primarily to encapsulate the fundamental alterations in supply and demand dynamics within financial markets. The premise of the OFI model is underpinned by the belief that an augmentation in the best bid or size at the best bid is indicative of an escalation in demand, while a diminution signifies a contraction in demand. Inversely, a decrease in the best ask or an expansion in the size at the best ask points to a surge in supply, while an increment in the best ask or a contraction in size at the best ask signals a decrease in supply. Consequently, this metric measures the change in the supply or demand, even at a high-frequency level.

In contrast, the Momentum Change metric, as proposed in this study, takes its conceptual basis from physical systems, and seeks to encapsulate the `momentum' of the order book system. It is calculated as the aggregated product of the velocity and the corresponding volume of each order. By incorporating the impact of each order on the price movement based on their LOB depth and quoted volume, this measure provides a more holistic view of the order book dynamics.

\subsection{Introduction on the OFI}

The OFI measure introduced by \cite{cont2014price} is a pivotal instrument in the domain of market microstructure, playing an instrumental role in assessing the swings in supply and demand dynamics. Conceived as an answer to the demand for a sophisticated understanding of these oscillations, the OFI measure is firmly rooted in empirical finance research and has been extensively employed as a tool for market examination.

In theoretical terms, the OFI measure is founded on the premise that the oscillations in the best bid and ask levels, in conjunction with the corresponding sizes, can act as an effective surrogate for the latent supply and demand changes in the market. In specific terms, an augmentation in the best bid or an augmentation in the size at the best bid is construed as a surge in demand, signaling an influx of purchasing interest in the market. Inversely, a diminution in these variables signifies a contraction in demand, indicating a shrinkage in buying interest.

On the supply facet, the OFI measure operates under an analogous rationale. A diminution in the best ask or an augmentation in the size at the best ask is perceived as a swell in supply, suggesting an increasing willingness of market participants to vend. In contrast, an augmentation in the best ask or a diminution in size at the best ask indicates a contraction in supply, signifying a retrenchment in selling interest.

The OFI measure, therefore, acts as a vital metric that encapsulates high-frequency changes in market dynamics. By providing real-time insights into the state of supply and demand in the market, the OFI measure aids the creation of reactive and adaptive trading strategies. It enables market participants to forecast price movements and adapt their positions accordingly, thus mitigating risk and optimizing returns.

\subsection{The Prediction Power on Price Change}

\begin{figure}
\begin{center}
\begin{minipage}{160mm}
\subfigure[BTC/USD market, 1-second change.]{
\resizebox*{8cm}{!}{\includegraphics{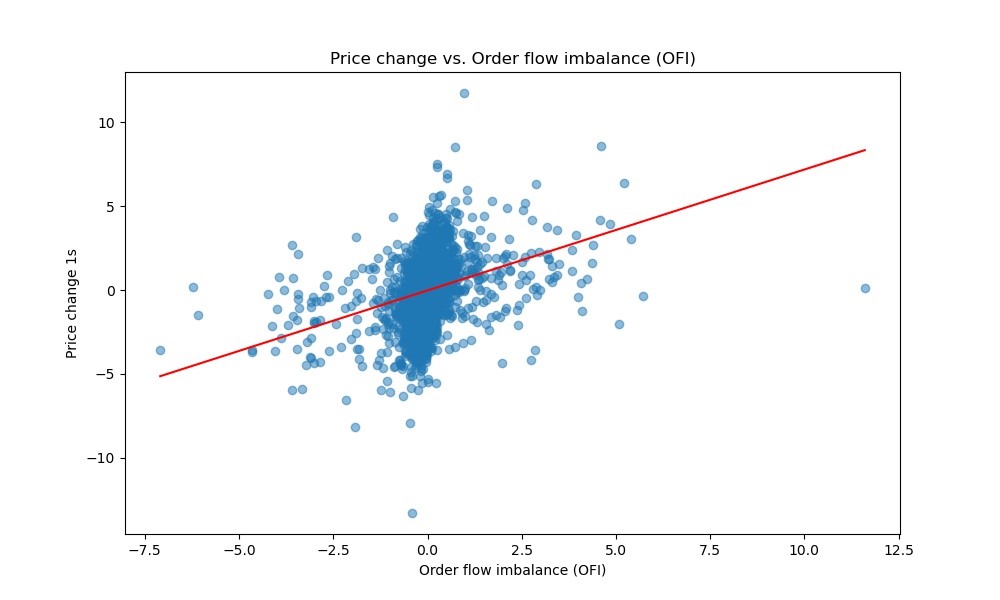}}\label{fig:BTC_OFI1}}
\subfigure[BTC/USD market, 10-second change]{
\resizebox*{8cm}{!}{\includegraphics{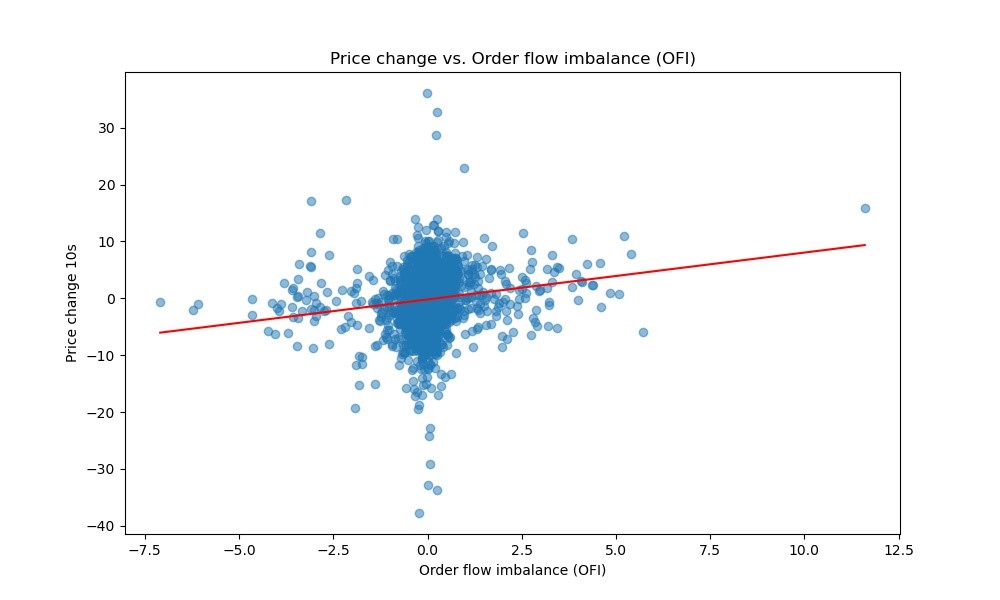}}\label{fig:BTC_OFI10}}
\caption{Regression analysis on the match price change against the OFI.
\label{fig:OFI}}
\end{minipage}
\end{center}
\end{figure}

\begin{figure}
\begin{center}
\begin{minipage}{160mm}
\subfigure[BTC/USD market, 1-second change.]{
\resizebox*{8cm}{!}{\includegraphics{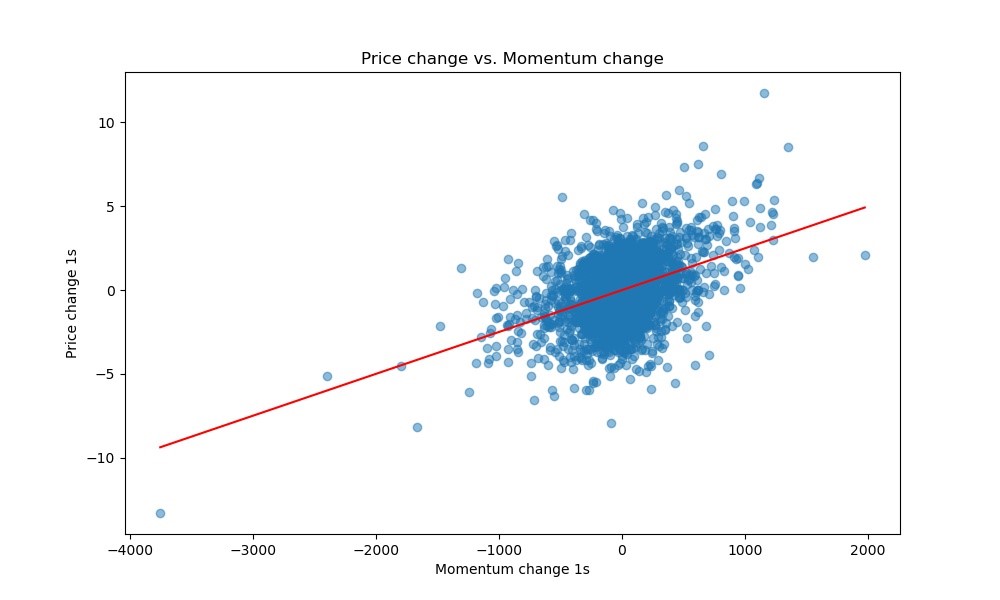}}\label{fig:BTC_momentum1}}
\subfigure[BTC/USD market, 10-second change]{
\resizebox*{8cm}{!}{\includegraphics{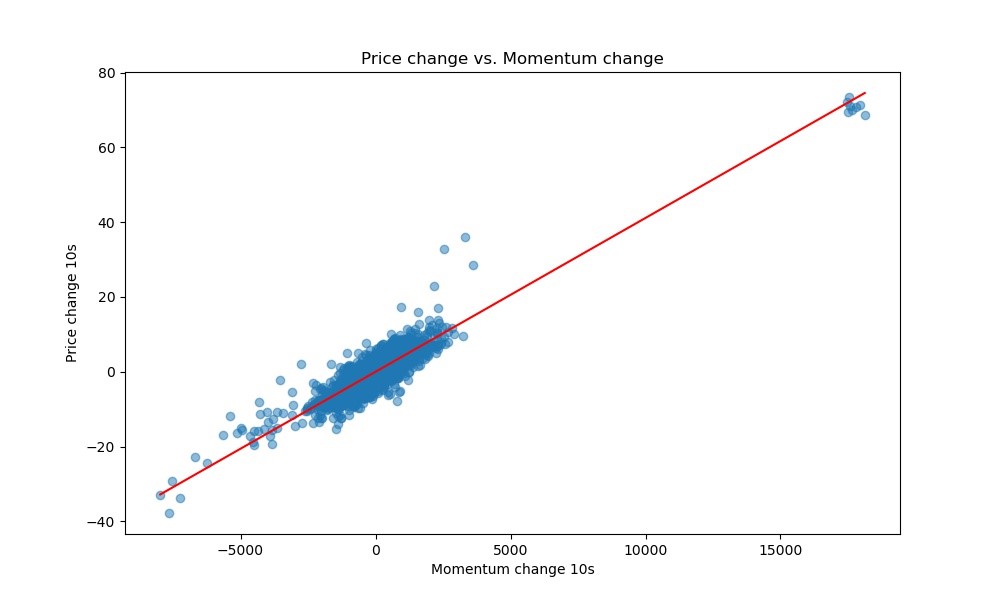}}\label{fig:BTC_momentum10}}
\caption{Regression analysis on the match price change against the momentum.
\label{fig:momentum}}
\end{minipage}
\end{center}
\end{figure}

\begin{figure}
\begin{center}
\begin{minipage}{160mm}
\subfigure[LUNA/USD market, 1-second change.]{
\resizebox*{8cm}{!}{\includegraphics{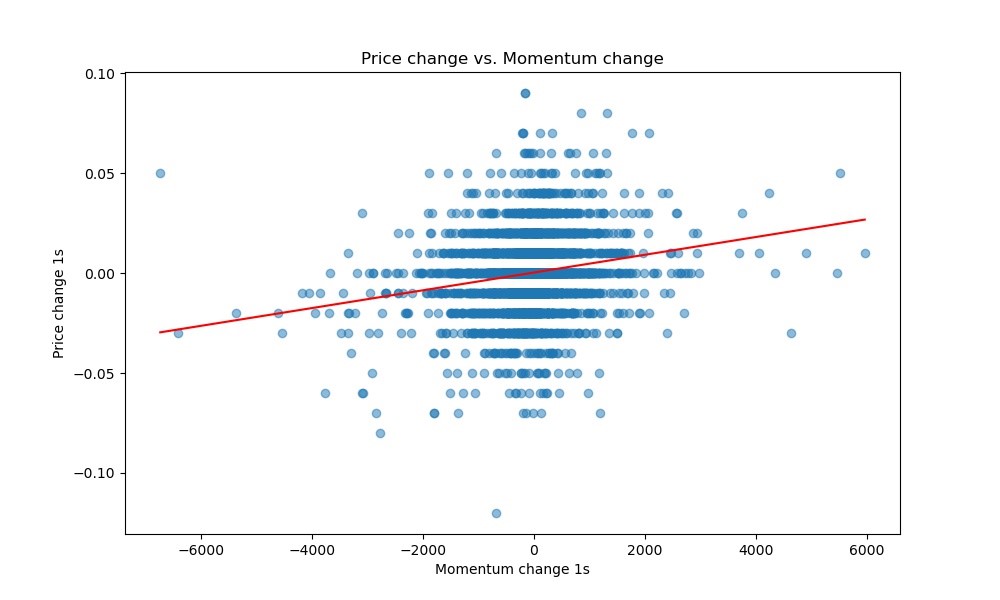}}\label{fig:LUNA_momentum1}}
\subfigure[LUNA/USD market, 10-second change]{
\resizebox*{8cm}{!}{\includegraphics{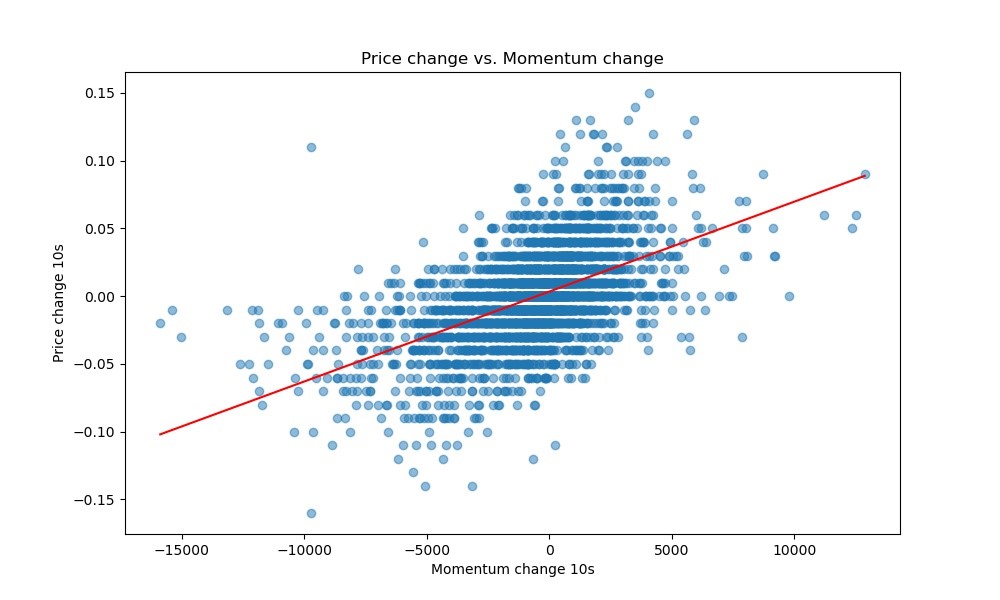}}\label{fig:LUNA_momentum10}}
\caption{Regression analysis on the match price change against the momentum.
\label{fig:momentum2}}
\end{minipage}
\end{center}
\end{figure}

This section provides a comprehensive empirical analysis of the predictive power of two different measures, namely the Order Flow Imbalance (OFI) and Momentum Change, in forecasting the price change of Bitcoin and LUNA over two distinct time horizons: one and ten seconds. For LUNA, due to unavailability of certain data points, the OFI measure could not be computed, and thus we only report the results for physical momentum change.

In order to evaluate the respective predictive abilities of trade imbalances and Order Flow Imbalances with regards to price fluctuations, we conduct a series of the following regressions:
\begin{equation}
    \Delta P_k = \alpha_i + \beta_i OFI_k + \epsilon_k
\end{equation}
for the OFI measure and 
\begin{equation}
    \Delta P_k = \alpha_i + \beta_i M_k + \epsilon_k
\end{equation}
for the momentum measure.

These regressions are computed independently for each data subset (indexed by $i$). Each subset represents an one-hour order book data sampled by 1-second frequency, which typically comprises approximately 3600 data points (indexed by $k$), collected at a frequency of one second. The details for these regression analysis are listed in Table~\ref{table:regression_analysis}.

The OFI measure was regressed against price changes of Bitcoin over one-second and ten-second horizons according to Figure~\ref{fig:OFI}. The R-squared values were 0.094 and 0.019 for the one-second and ten-second horizons, respectively. These relatively low R-squared values suggest that the predictive power of the OFI measure is modest, but not negligible. The F-statistic results (373.0 and 70.87) and their corresponding p-values (3.83e-79 and 5.45e-17) indicate that the OFI measure is statistically significant in predicting Bitcoin price changes. The coefficients of the OFI measure were positive and significant at the 0.000 level, implying that an increase in OFI leads to an increase in Bitcoin price. The performance of the OFI on Bitcoin data is less statistically significant than that on NYSE TAQ stock data with R-squared value of about 0.65 \cite{cont2014price}.

The Momentum Change measure was regressed against price changes of Bitcoin and LUNA over one-second and ten-second horizons in Figure~\ref{fig:momentum}. For Bitcoin, the R-squared values were 0.168 and 0.816 for the one-second and ten-second horizons, respectively, suggesting a stronger predictive power than the OFI measure. The F-statistics (723.6 and 1.589e+04) and their corresponding p-values (2.33e-145 and 0.00) further support the significant role of Momentum Change in predicting Bitcoin price changes. The coefficients of the Momentum Change measure were positive and significant at the 0.000 level, suggesting that an increase in momentum would lead to a price increase.

For LUNA, the R-squared values were 0.036 and 0.274 for the one-second and ten-second horizons respectively (Figure~\ref{fig:momentum2}). This demonstrates that Momentum Change has a notable predictive power for LUNA prices, especially over a ten-second horizon. Although the performance is not as good as that for Bitcoin, but it is better than using the OFI measure. The F-statistics (137.0 and 1375) and their corresponding p-values (4.26e-31 and 1.43e-255) confirm the statistical significance of Momentum Change in predicting LUNA price changes. The coefficients of the Momentum Change measure were positive and significant at the 0.000 level, implying that an increase in momentum leads to a price increase.

Overall, our empirical analysis reveals that the Momentum Change measure has a stronger predictive ability for the price change of both Bitcoin and LUNA, compared to the OFI measure for Bitcoin. This suggests that the Momentum Change measure could be a more robust tool for short-term price prediction in the cryptocurrency market.

\begin{table}
\begin{center}
    \caption{The regression analysis}
    \label{table:regression_analysis}
    \resizebox{\textwidth}{!}{
    \begin{tabular}{lrrrrrrrrrr}
        \hline
        \rule{0pt}{6pt}
        Dataset & $R^2$ & F-statistic & Prob(F-statistic) & Omnibus  & Prob(Omnibus) & Skew & Kurtosis & Durbin-Watson & Jarque-Bera (JB) & Prob(JB)\\ 
        \hline
        \\ [-6pt]
        BTC/USD OFI vs $\Delta P$ 1s & 0.094 & 373.0 & 3.83e-79 & 255.040 & 0.000 & -0.139 & 5.829 & 2.472 & 1208.261 & 4.26e-263\\
        BTC/USD OFI vs $\Delta P$ 10s & 0.019 & 70.87 & 5.45e-17 & 655.211 & 0.000 & -0.403 & 11.230 & 0.337 & 10204.282 & 0.00\\
        BTC/USD $\Delta M$ vs $\Delta P$ 1s & 0.168 & 723.6 & 2.33e-145 & 110.277 & 0.000 & -0.128 & 4.296 & 2.653 & 261.177 & 1.93e-57\\
        BTC/USD $\Delta M$ vs $\Delta P$ 10s & 0.816 & 1.589e+04 & 0.00 & 747.131 & 0.000 & 0.737 & 9.455 & 0.970 & 6560.394 & 0.00\\
        LUNA/USD $\Delta M$ vs $\Delta P$ 1s & 0.036 & 137.0 & 4.26e-31 & 202.154 & 0.000 & 0.012 & 5.316 & 2.793 & 816.435 & 5.17e-178\\
        LUNA/USD $\Delta M$ vs $\Delta P$ 10s & 0.274 & 1375 & 1.43e-255 & 165.117 & 0.000 & 0.236 & 4.593 & 0.840 & 419.270 & 9.05e-92\\
        \hline
    \end{tabular}}
\end{center}
\end{table}

\begin{table}
\begin{center}
    \caption{The fitted model}
    \label{table:fitted_model}

    \begin{tabular}{lrrrrrr}
        \hline
        \rule{0pt}{6pt}
        The fitted model & Coefficient & Standard error & $t$ & $P\ge |t| $ & [2.5\%, & 97.5\%] \\ 
        \hline
        \\ [-6pt]
        $\alpha_1$ & -0.0215 & 0.027 & -0.787 & 0.431 & -0.075 & 0.032\\
        $\beta_1$ & 0.7208 & 0.037 & 19.313 & 0.000 & 0.648 & 0.794\\
        \hline
        $\alpha_2$ & -0.2008 & 0.072 & -2.803 & 0.005 & -0.341 & -0.060\\
        $\beta_2$ & 0.8237 & 0.098 & 8.418 & 0.000 & 0.632 & 1.016\\
        \hline
        $\alpha_3$ & -0.0070 & 0.026 & -0.269 & 0.788 & -0.058 & 0.044\\
        $\beta_3$ & 0.0025 & 9.28e-05 & 26.900 & 0.000 & 0.002 & 0.003\\
        \hline
        $\alpha_4$ & 0.0005 & 0.039 & 0.013 & 0.989 & -0.076 & 0.077\\
        $\beta_4$ & 0.0041 & 3.26e-05 & 126.061 & 0.000 & 0.004 & 0.004\\
        \hline
        $\alpha_5$ & 0.0002 & 0.000 & 0.729 & 0.466 & -0.000 & 0.001\\
        $\beta_5$ & 4.441e-06 & 3.79e-07 & 11.706 & 0.000 & 3.7e-06 & 5.18e-06\\
        \hline
        $\alpha_6$ & 0.0034 & 0.000 & 7.305 & 0.000 & 0.003 & 0.004\\
        $\beta_6$ & 6.637e-06 & 1.79e-07 & 37.075 & 0.000 & 6.29e-06 & 6.99e-06\\
        \hline
    \end{tabular}
\end{center}
\end{table}

\subsection{Comparative Analysis}

The empirical study offers a comparative evaluation of these two metrics. For Bitcoin, it was discovered that the Momentum Change metric demonstrated superior performance than the OFI in forecasting price fluctuations over one-second and ten-second intervals. The R-squared values associated with the Momentum Change metric were significantly larger than those affiliated with the OFI metric, suggesting a more potent predictive capacity. The statistical significance of the Momentum Change metric, as evidenced by the F-statistics and their corresponding p-values, further accentuates its efficacy in forecasting Bitcoin price fluctuations.

In the case of LUNA, due to data limitations, only the Momentum Change metric could be computed. However, the results indicated a substantial predictive capability of the Momentum Change metric for LUNA prices, especially over a ten-second interval.

The results imply that while the OFI metric provides critical insights into the dynamics of supply and demand, the Momentum Change metric, by amalgamating price and volume fluctuations, offers a more comprehensive understanding of market dynamics. This underscores the potential advantages of integrating concepts from the physical sciences into financial market analysis. The superior performance of the momentum measure over the OFI implies potential limitations of the OFI in forecasting price fluctuations in cryptocurrency data, particularly when compared with stock data. This underscores the distinct characteristics of cryptocurrency markets and the necessity for further experiments of the physics-based model on the other asset classes.

The comparative empirical study between the Momentum Change metric and the OFI model constitutes a notable addition to the literature on price change prediction. It not only introduces a fresh perspective but also indicates a paradigm shift from traditional econometric models to an interdisciplinary approach that merges financial economics and physical sciences. Further investigation into this interdisciplinary approach could potentially facilitate advancements in price change prediction within financial markets.

\section{Momentum and Deep LSTM Model on Price Prediction}
\label{Benchmark}

Based on the work in \cite{sirignano2019deep}, the Deep Long Short Term Memory (LSTM) model designed by \cite{sirignano2019universal} was constituted by three layers of LSTM units, succeeded by a fully-connected layer of rectified linear units (ReLUs). The subsequent application of a softmax activation function facilitated the generation of a probability distribution for the future price movement. \cite{sirignano2019universal} compared the performance of the deep LSTM model against the other linear models, especially Vector Autoregressive (VAR) model. The empirical results revealed that the deep LSTM model transcended these counterparts, demonstrating its capability to capture non-linear relationships in the LOB data which other models might miss.\noindent

\subsection{Introduction on the deep LSTM model}

\cite{sirignano2019universal} utilised a deep feedforward neural network, where the input was several layers of the limit order book (i.e., levels of bids and asks, quantities at each level, etc.), and the output was the future price change after a certain fixed time horizon. They used a large dataset to train the neural network, optimising the model's weights using backpropagation. The model was trained to minimise the difference between its predicted price changes and the actual price changes observed in the data.

Instead of handcrafting features as in traditional finance models, the deep LSTM model was trained to directly learn from the order book data. This automatic feature extraction capability is crucial for understanding complex structures and patterns in financial data. 

\cite{sirignano2019universal} compared the performance of the deep LSTM model with other benchmark models. The deep LSTM outperformed these models, demonstrating its capability to capture non-linear relationships in the LOB data which other models might miss. Apart from its predictive power, the model's importance lies in its ability to shed light on the microstructure of financial markets. The patterns and features automatically learned by the deep LSTM model can be interpreted to gain insights into the underlying mechanisms of price formation in financial markets.

In essence, their study showcases the power of deep learning in capturing and understanding the complex dynamics of financial markets, particularly when it comes to the microstructure of how prices are formed. It also emphasises the potential of deep learning to unveil universal features of price formation.

\subsection{Implementation of the deep LSTM model}

This subsection aims to provide a clear and academically rigorous overview of the LSTM model's architecture and parameters implemented according to the \cite{sirignano2019deep} and \cite{sirignano2019universal}.

\begin{figure}[!ht]
\begin{center}
\includegraphics[width=\textwidth]{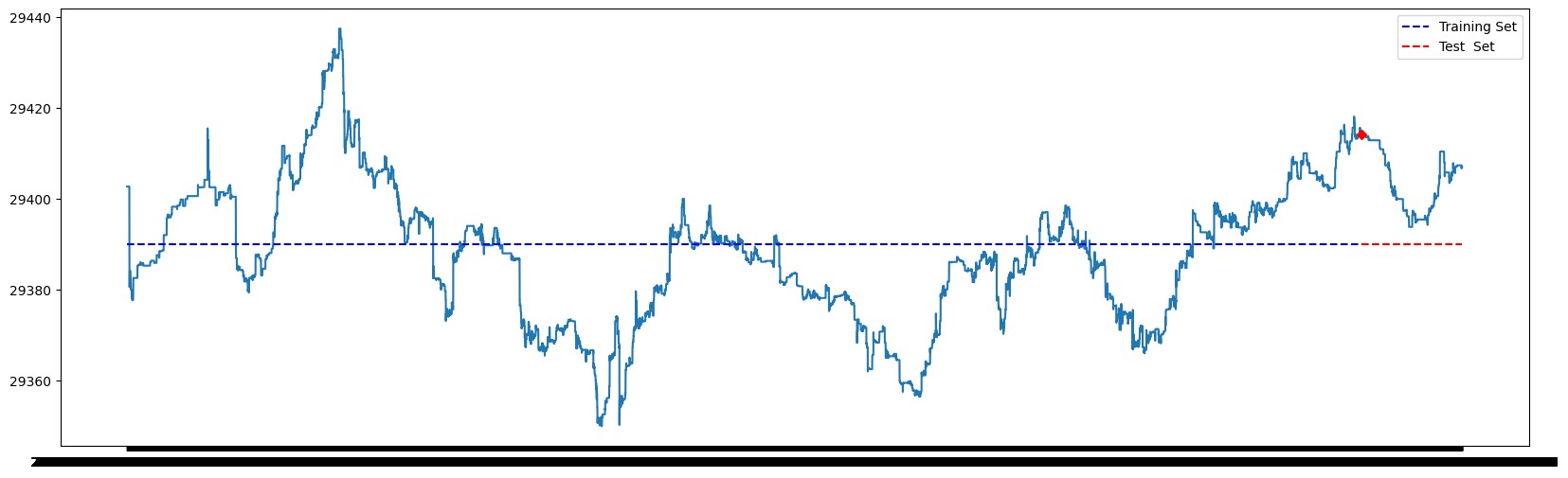}
\end{center}
\caption{Measures on LUNA/USD market}
\label{fig:lstm_dataset}
\end{figure}

Our dataset comprises snapshots of the LOB, encapsulating price and volume data for the 20 highest bid and 20 lowest ask orders. The snapshots data was reconstructed from the Coinbase exchange for Bitcoin/USD pair during 2023/08/11 23:00 - 2023/08/12 12:00. Note that we did not test the depp LSTM model on LUNA/USD pair during flash crash due to the lack of corresponding snapshots data. Each observation or timestamp thus constitutes 80 features: 20 bid prices, 20 bid volumes, 20 ask prices, and 20 ask volumes. The dataset is contained with 44215 snapshots of the LOB in the span of 13 hours, and the 40889 of them are used as trained set while the left for the test set. The classification of training set and test set is depicted in Figure~\ref{fig:lstm_dataset}, where the data labelled with blue dotted line is the training set and red for the test set.

Before feeding the data into the LSTM, we subjected it to rigorous preprocessing. We standardized the dataset ensuring zero mean and unit variance for each feature. This scaling is imperative for neural networks to ensure all input features are treated with equal importance and to aid in model convergence.

The architecture of the LSTM model is as follows:

Input layer: The model is designed to accommodate sequences of variable length, each embedding 80 features, reflective of the bid-ask dynamics in our dataset.

\begin{figure}[!ht]
\begin{center}
\includegraphics[height=2.5in]{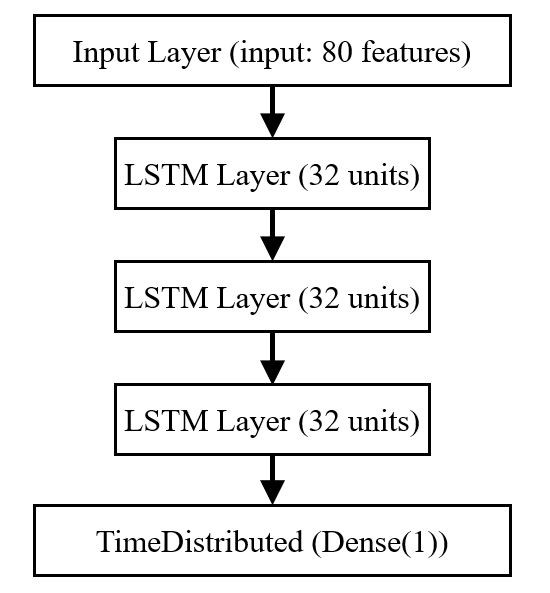}
\end{center}
\caption{Diagram on the architecture of the deep LSTM model}
\label{fig:lstm_diagram}
\end{figure}

Deep LSTM layers: The subsequent layers are 3 LSTM layers each composed of 32 units. The activation function chosen is the hyperbolic tangent (tanh), renowned for its ability to manage vanishing gradients, especially in deeper networks. The recurrent activations utilise the hard-sigmoid function, which is a deliberate configuration optimising for gradient flow and computational efficiency. The choice of initialisers for the kernels (Glorot Uniform) and recurrent weights (Orthogonal) are aligned with best practices to ensure effective weight initialization, which can profoundly impact the training dynamics. The subsequent two LSTM layers, mirroring their predecessor in units and configurations, offer another two layers of abstraction, further enhancing the model's capacity to discern intricate patterns within the data. The structure of the deep LSTM layers are demonstrated in Figure~\ref{fig:lstm_diagram}.

Output layer: Post the LSTM layers, the model employs a TimeDistributed wrapper around a dense layer with a linear activation function. This configuration enables the model to make predictions for each time step independently, providing a continuous output corresponding to the asset's predicted price.

Model Parameters:
The chosen optimiser for the model was RMSprop, favored for its adaptability in adjusting learning rates during training. The loss function, mean squared error (MSE), was selected given the regression nature of the task. The primary metric for model evaluation was set as accuracy.

The model was trained over 75 epochs. The batch size, a crucial hyperparameter, was empirically set at 30. This choice signifies that the model processes 30 sequences simultaneously in each iteration, balancing computational efficiency with gradient estimation accuracy. Such a model aims to offer enhanced predictive capabilities, capturing the nuanced temporal patterns that drive price dynamics.

\subsection{The Prediction accuracy}

The model performance were estimated using the prediction accuracy function given by \cite{sirignano2019universal} in equation \ref{eq:accuracy}, as their models were trained to forecast the direction of the next price change. Note that the number of correct prediction is calculated from the total correct forecasts of the model on the direction of price changes every time an event of price change occurs, where the datapoints are not sampled equally but are event-driven.

\begin{equation}
    Accuracy = \frac{\texttt{Number of correct predictions}}{\texttt{Total number of price changes}} \times 100\%.
\label{eq:accuracy}
\end{equation}

\cite{sirignano2019universal} reported a prediction accuracy of approximately 57\%-67\% for stock-specific linear (VAR) models, and 65\%-75\% for deep LSTM model on 500 NASDAQ stocks in total.

\begin{figure}[!ht]
\begin{center}
\includegraphics[width=\textwidth]{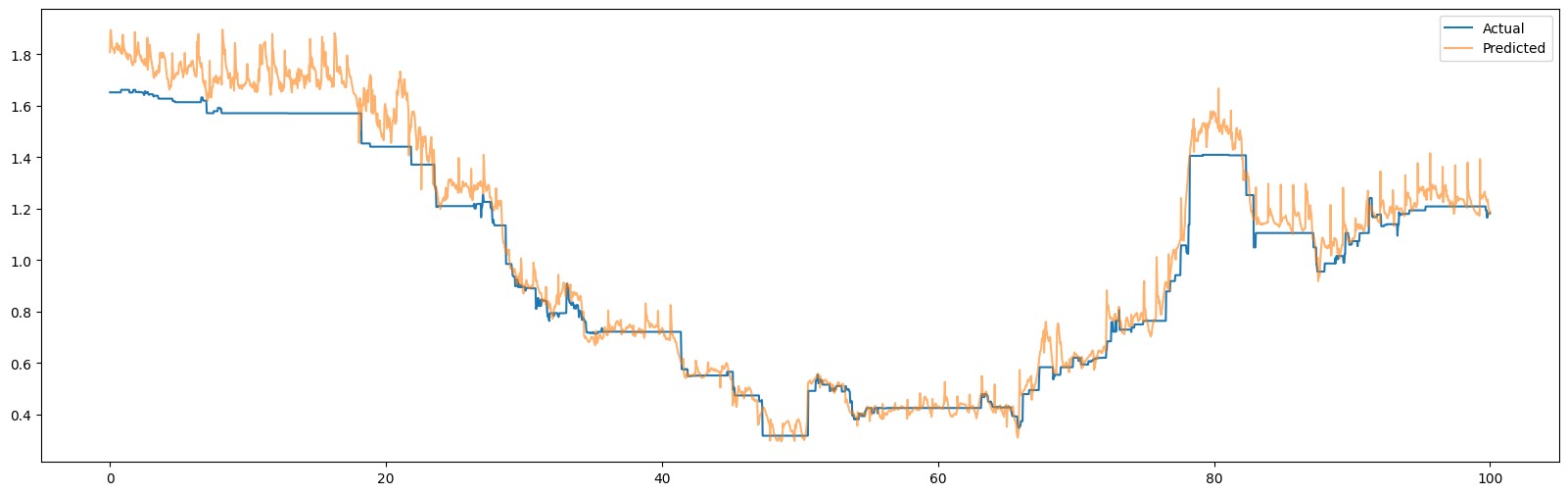}
\end{center}
\caption{Prediction performance of deep LSTM model on the 1-second price change}
\label{fig:lstm_1s}
\end{figure}

\begin{figure}[!ht]
\begin{center}
\includegraphics[width=\textwidth]{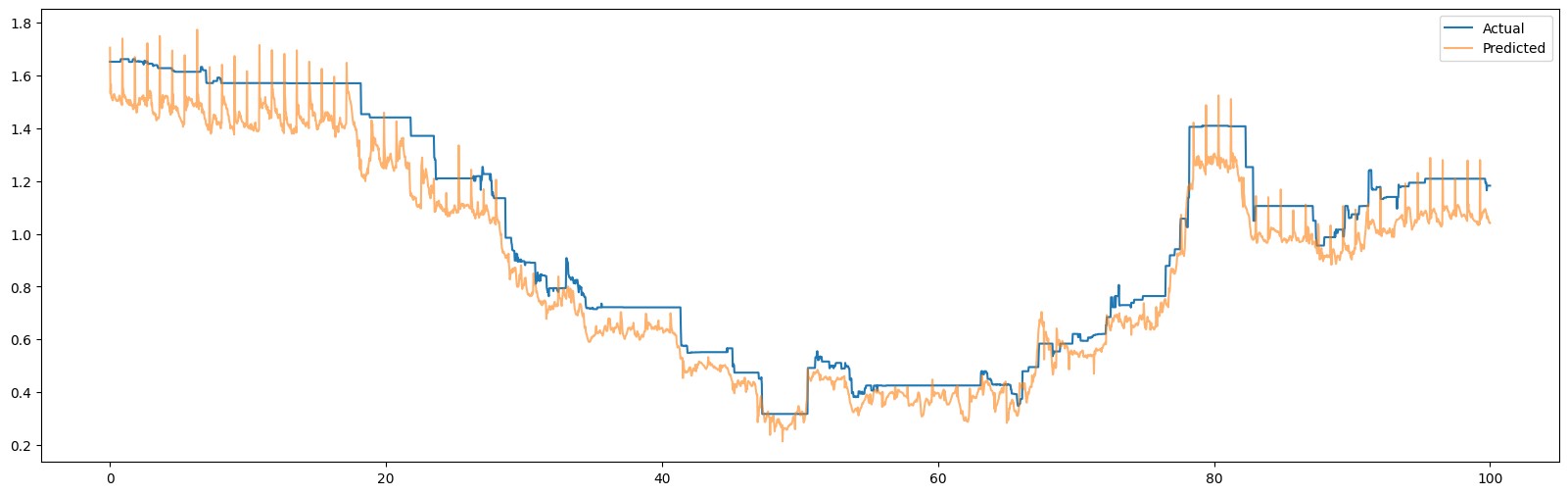}
\end{center}
\caption{Prediction performance of deep LSTM model on the 10-second price change}
\label{fig:lstm_10s}
\end{figure}

We trained their deep LSTM model for the Bitcoin/USD market using the 12-hour training set of LOB snapshots, then the model was inputed with the LOB snapshots in the 1-hour test set to predict the price changes 1 second and 10 seconds later respectively (the performances of each prediction are given in Figure~\ref{fig:lstm_1s} and \ref{fig:lstm_10s} separately. Meanwhile, the physics-based model were deployed on the Level 3 order book data during the same period of the 1-hour test set. The prediction accuracy was about 60.12\% for 1-second price change and 62.57\% for 10-second change. In contrary, the prediction accuracy of the physics-based model is 60.25\% for 1-second price change and 78.69\% for 10-second on the BTC/USD pair. In addition, even for the LUNA/USD dataset during flash crash which is harder to predict, the physics-based model still reaches an relative high accuracy of 58.70\% and 69.82\%.

The decrease in the accuracy of the deep LSTM model on cryptocurrency market being compared to the stock market may be resulted from the difference of the market microstructure as the the stock market has better liquidity and the LOB considered as more dense, while the bid-ask quotes in the LOB spread on sparse depths. In other words, 20 bid and ask quotes of the LOB in cryptocurrency market contain much less information than the stock market. However, the deep learning method is not feasible to process the data with more depth due to the computing power and model complexity restrictions.

\begin{table}[!ht]
\begin{center}
    \caption{The comparison on prediction accuracy}
    \label{table:accuracy}
    \begin{tabular}{lrrr}
        \hline
        \rule{0pt}{6pt}
        Accuracy in \%                  & 1-second price change & 10-second price change \\
        \\ 
        \hline
        \\ [-6pt]
        linear VAR model on stocks      & 57-67                 & N/A                    \\
        \hline
        deep LSTM model on stocks       & 65-75                 & N/A                    \\
        \hline
        deep LSTM model on BTC/USD      & 60.12                 & 62.57                  \\
        \hline
        physics-based model on BTC/USD  & 60.25                 & 78.69                  \\
        \hline
        physics-based model on LUNA/USD & 58.70                 & 69.82                  \\
        \hline
    \end{tabular}
\end{center}
\end{table}

\subsection{Comparative Analysis}

{\bf Data dependencies and computational constraints.} The performance of deep learning models heavily relies on the quality and quantity of the data. High-quality, expansive data sets are paramount for robust model training, ensuring generalization and mitigating overfitting risks. Furthermore, the computational burden of these models, especially when delving into extensive datasets, necessitates sophisticated hardware, such as GPUs, making the method less accessible for researchers or practitioners without the necessary resources. Specifically, \cite{sirignano2019deep} and \cite{sirignano2019universal} used 25-50 GPUs to train their deep LSTM model on nearly 500 stocks for the time period spanning over a 17-month span, as such a dataset contains 50 terabytes of data. Both the quantity of the data and the number of GPUs demanded constrain the accessibility of this method. Conversely, the physics-based model, devoid of training data dependency, increases the practicability as well as avoids possible delay caused by the model complexity, hence is capable of extending to further trading strategies and broader applications.

{\bf Model Interpretability and Nonlinear Representations.} \cite{sirignano2019universal} claimed that their deep LSTM model's superiority over other linear models was attributed to the model's ability on representing the nonlinear relationships between the price dynamics and the order book that resulted from the market supply and demand. Their assertions align with a rich strand of empirical and econometric studies on financial markets. However, one of the primary challenges with deep learning models is their lack of interpretability. These models are often labeled as `black boxes' because of the difficulty to understand the precise reasoning behind their predictions. While the deep learning can identify patterns and make predictions, understanding why they make those predictions is less straightforward. In juxtaposition, the physics-based model pioneers an intuitive technique to gauge such nonlinear relationships in the LOB modelled as complex system in a straightforward way. In contrast to the large amount of parameters contained in the deep learning models, the simplicity of this model further eliminates the chance of parameter overfitting.

{\bf Order Book Depth and Price Dynamics.} The study of \cite{cont2014price} discovered that while the first level has the strongest impact on price dynamics, deeper depths still contain information relevant for forecasting. The experimental findings from \cite{sirignano2019universal} further underscore and prove the predictive significance of the deeper depths in the LOB on price dynamics. They defined and calculated sensitivity of a given depth in the LOB, which was 1.7\% for the best bid/best ask, 0.4\% for depth 2, 0.2\% for both depth 3 and 4, and the total sensitivity of depth 5-10 was 0.4\%. The truth is the depth beyond contain even more significant information, however, the restriction on computing power and model complexity of deep learning models make it difficult to train with the data for deeper depth. Contrarily, the physics-based model decides the depths relevant to the price dynamics by calculating the defined active depth in order to take full advantage. As a result, the physics-based model would be more efficient to capture the price dynamics compared to deep learning models.

\section{Conclusion}
\label{Conclusion}

We conducted an analysis of Level 3 order book data rather than relying on time-based snapshots, allowing us to study the market microstructure in the microscopic scale while retaining all temporal information.
We modeled the orders as physical particles, and developed the concept of active area to calculate the momentum of each order and gauge its impact on the order book. This enabled us to create a systematic description of the state of the order book, using a vast amount of Level 3 data in an efficient manner. We then studied the market behaviour and found that limit and market orders had opposite patterns of behaviour during the flash crash of the LUNA cryptocurrency.

Our research approach utilised Level 3 order book data, enabling us to examine the market microstructure with a high degree of temporal resolution. Rather than adopting time-based snapshots, our analysis preserved the complete temporal context. In this framework, orders were conceptualized as physical particles, and we introduced the notions of active area to calculate the kinetic energy and momentum associated with each order and assess their impact on the order book. This approach facilitated a systematic depiction of the order book state, harnessing the extensive amount of Level 3 data in an effective manner. Upon analysing the market behaviour, we discovered contrasting patterns in the activities of limit and market orders during the flash crash incident involving the LUNA cryptocurrency.

Moreover, the kinetic energy and momentum measures offer promising predictive capabilities for price volatility and impact in the context of cryptocurrency markets. Empirical findings suggest that these new measures outperform established models including VPIN, OFI, and deep LSTM model, highlighting the potential of interdisciplinary approaches in financial market prediction. The physics-based model's capacity to harness deeper depths in the LOB, vital for forecasting, without computational hindrances further accentuates its advantages, and provides a more efficient, interpretable, and adaptable solution for understanding market dynamics. This research opens avenues for further exploration of the applicability of physical principles in financial markets and encourages future studies to continue bridging the gap between these two distinct disciplines.

The method presented in this article has the potential to inspire future studies that use statistical physics concepts to understand financial markets. One direction for future work could be to incorporate other physical concepts, while another interesting avenue could be to analyse the other asset classes, such as equities and futures.

\section*{Acknowledgments}
We would like to thank the valuable discussions with Adam Ostaszewski, Merlin Mei, Ke Chen, and Yue Xiao.

\bibliography{rQUFguide}
\end{document}